\documentclass[reprint, superscriptaddress,amsmath,amssymb,aps,prb,]{revtex4-1}

\pdfoutput=1

\usepackage{graphicx}
\usepackage{dcolumn}
\usepackage{bm}
\usepackage{verbatim}
\usepackage{color}
\usepackage{hyperref}

\newcommand{\ket}[1]{\vert #1 \rangle}

\begin{document}

\title{Photonic Simulation of Entanglement Growth After a Spin Chain Quench}

\author{Ioannis Pitsios}
\affiliation{Istituto di Fotonica e Nanotecnologie - Consiglio Nazionale delle Ricerche (IFN-CNR), P.za Leonardo da Vinci, 32, I-20133 Milano, Italy}
\affiliation{Dipartimento di Fisica - Politecnico di Milano, P.za Leonardo da Vinci, 32, I-20133 Milano, Italy}

\author{Leonardo Banchi}
\affiliation{Department of Physics and Astronomy, University College London, Gower Street, WC1E 6BT London, United Kingdom}

\author{Adil S. Rab}
\affiliation{Dipartimento di Fisica - Sapienza Universit\`{a} di Roma, P.le Aldo Moro 5, I-00185 Roma, Italy}

\author{Marco Bentivegna}
\affiliation{Dipartimento di Fisica - Sapienza Universit\`{a} di Roma, P.le Aldo Moro 5, I-00185 Roma, Italy}

\author{Debora Caprara}
\affiliation{Dipartimento di Fisica - Sapienza Universit\`{a} di Roma, P.le Aldo Moro 5, I-00185 Roma, Italy}

\author{Andrea Crespi}
\affiliation{Istituto di Fotonica e Nanotecnologie - Consiglio Nazionale delle Ricerche (IFN-CNR), P.za Leonardo da Vinci, 32, I-20133 Milano, Italy}
\affiliation{Dipartimento di Fisica - Politecnico di Milano, P.za Leonardo da Vinci, 32, I-20133 Milano, Italy}

\author{Nicol\`o Spagnolo}
\affiliation{Dipartimento di Fisica - Sapienza Universit\`{a} di Roma, P.le Aldo Moro 5, I-00185 Roma, Italy}

\author{Sougato Bose}
\email{s.bose@ucl.ac.uk}
\affiliation{Department of Physics and Astronomy, University College London, Gower Street, WC1E 6BT London, United Kingdom}

\author{Paolo Mataloni}
\affiliation{Dipartimento di Fisica - Sapienza Universit\`{a} di Roma, P.le Aldo Moro 5, I-00185 Roma, Italy}

\author{Roberto Osellame}
\email{roberto.osellame@polimi.it}
\affiliation{Istituto di Fotonica e Nanotecnologie - Consiglio Nazionale delle Ricerche (IFN-CNR), P.za Leonardo da Vinci, 32, I-20133 Milano, Italy}
\affiliation{Dipartimento di Fisica - Politecnico di Milano, P.za Leonardo da Vinci, 32, I-20133 Milano, Italy}

\author{Fabio Sciarrino}
\email{fabio.sciarrino@uniroma1.it}
\affiliation{Dipartimento di Fisica - Sapienza Universit\`{a} di Roma, P.le Aldo Moro 5, I-00185 Roma, Italy}

\begin{abstract}
{\bf The time evolution of quantum many-body systems is one of the least understood frontiers of physics. The most curious feature of such dynamics is, generically, the growth of quantum entanglement with time to an amount proportional to the system size (volume law) even when the interactions are local. This phenomenon, unobserved to date,  has great ramifications for fundamental issues such as thermalisation and black-hole formation, while its optimisation clearly has an impact on technology (e.g., for on-chip quantum networking). Here we use an integrated photonic chip to simulate the dynamics of a spin chain, a canonical many-body system. A digital approach is used to engineer the evolution so as to maximise the generation of entanglement. The resulting volume law growth of entanglement is certified by constructing a second chip, which simultaneously measures the entanglement between multiple distant pairs of simulated spins. This is the first experimental verification of the volume law and opens up the use of photonic circuits as a unique tool for the optimisation of quantum devices. }
\end{abstract}

\maketitle

The universal features of quantum many-body dynamics remain poorly understood sparing the glimpses provided by a small set of solvable models \cite{sengupta}. This is largely because a classical computer fails to keep track of the extensive growth of quantum entanglement in typical cases. One way for enhancing our understanding of such dynamics is to mimic it with another, more controllable ``quantum" system -- a quantum simulator, as suggested by Feynman \cite{Feynman-simulating}. It is then imperative that a true quantum simulator platform, which eventually aims to outperform classical computers, should be able to capture the extensive growth of entanglement in a many-body simulation. For example, a sudden global change of a many-body Hamiltonian (a quantum quench) can induce a dynamics that ultimately entangles two complementary parts of the system by an amount  proportional to their size (a volume law) \cite{Calabrese-Cardy, Altman-review, Moore, Banchi-Bose}. This quench induced entanglement growth is a phenomenon of pivotal importance, as it not only underpins the ubiquitous process of thermalisation \cite{Altman-review, Jens-review}, but, through a holographic correspondence, the dynamical formation of black holes in space-time \cite{Takayanagi, Liu-PRL-Tsunami}. Despite the recent remarkable measurements of entanglement \cite{Islam15} and its ``propagation" \cite{Jurcevic14, Richerme14, Fukuhara15} due to quenches in atomic/ionic many-body simulators, the key feature of entanglement growth in proportion to system size remains unverified. Although this verification, for fundamental reasons, is surely one of the next big aims of atomic/ionic simulators as mentioned in Ref.\cite{Islam15}, the feat is, nonetheless, yet to be achieved.  

Verification of the volume law is not only of fundamental importance. If the entanglement growth can be controlled and optimised, it could unlock the potential of many-body systems as on-chip entanglers for quantum networking \cite{Banchi-Bose,Wichterich09}. The same high degree of control also enables its usage as ballistic quantum wires \cite{Bose03,Christandl04,Banchi10,Yao-Lukin,Banchi11} and computers \cite{Feynman-computer-optics}. This degree of control in entanglement generation is yet to be achieved in any quantum simulator. 

Photonic circuits have been used for realising varied phenomena such as bosonic and fermionic quantum walks \cite{Sansoni12,Matthews13}, quantum-classical differences in complexity \cite{boson-sampling,boson-sampling2,boson-sampling3,boson-sampling4}, quantum chemistry \cite{chem} and localization in transport \cite{CrespiAnderson}. The ground states of some few-body spin systems \cite{verstraete, walther-frustration, mataloni}, as well as some analogues of
quantum state transfer (QST) through spin chains using coherent optics \cite{bellec-nikolopoulos,alastair}, have been realised in photonics.
In this article, we report two firsts in the general area of quantum simulation of many-body dynamics. Using the photonic circuits we (i) verify the generation of extensive (volume law) entanglement from the dynamics of a spin chain, a canonical many-body system, following a quench, and (ii) demonstrate an exquisite control and optimisation of the generated entanglement, as necessary for building future devices.

\begin{figure*}[ht!]
    \includegraphics[width=0.78\textwidth]{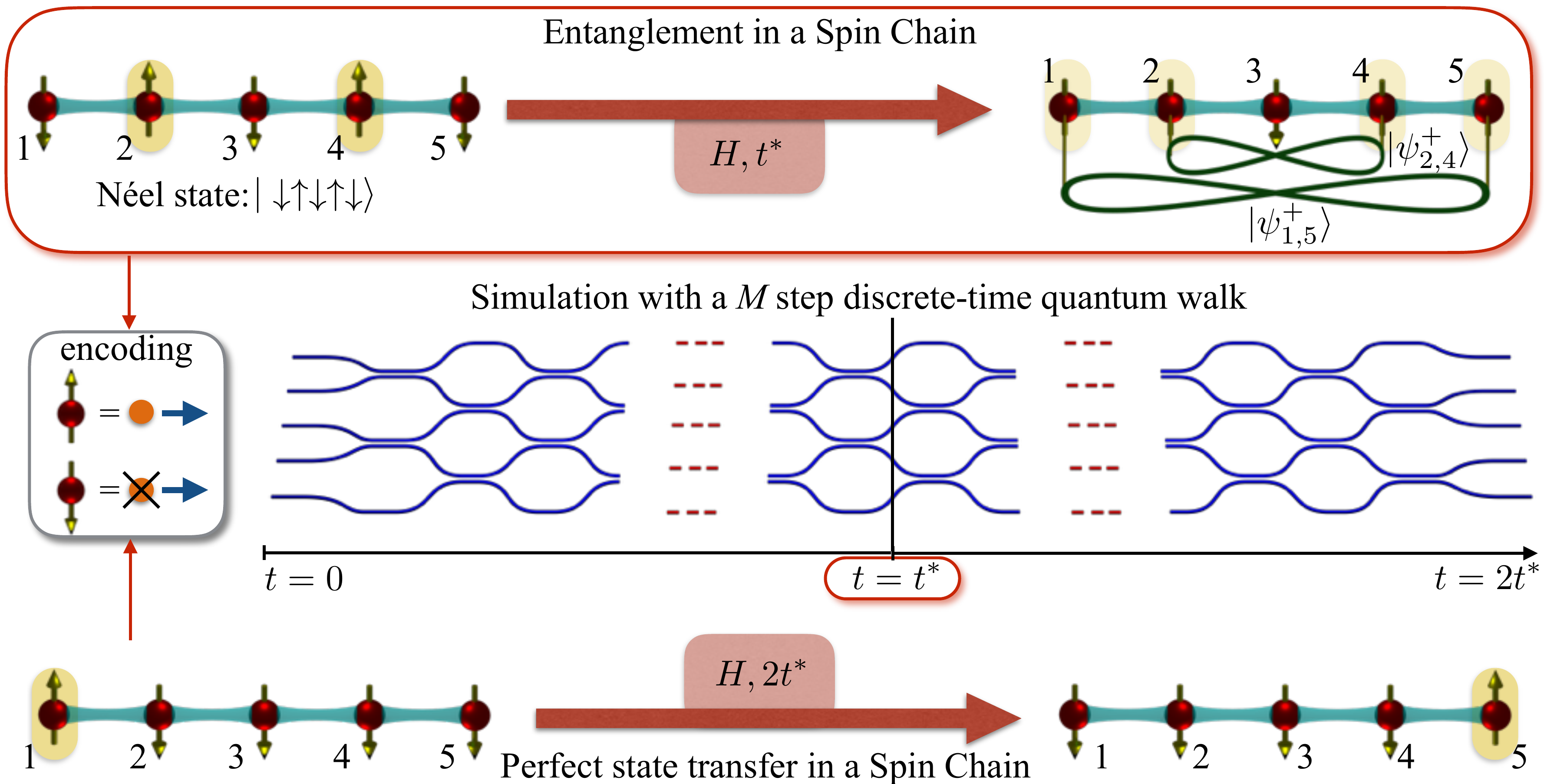}
 \caption{Quantum simulation of spin chain dynamics and the entanglement growth therein in photonic platforms. Fermions simulated on the network map, via the Jordan-Wigner transformation, to spin excitations in a chain: occupied and unoccupied modes correspond respectively to spin up and spin down.  The couplings of the effective spin chain Hamiltonian are so engineered as to generate, starting from a N\'{e}el state, at a time $t^{*}$, a state with maximal entanglement between distant symmetric sites with respect to the center of the chain. $|\psi^{+}_{i,j}\rangle$ stands for a maximally entangled state of spins in sites $i$ and $j$. This rainbow state exhibits a volume law entanglement in which the entanglement between the left and the right halves of the chain is $\sim N/2$. The time evolution is approximated through a set of discrete steps (a digital approach). If the dynamics was continued up to time 2$t^{*}$, it would implement an approximately perfect QST.} \label{concept}
\end{figure*}

Our achievements are made possible by engineering two novel photonic chips: (a) the quantum transport chip and (b) the entanglement characterization chip. The first chip is designed to implement a spin chain dynamics which generates, multiple entangled pairs of distant ``simulated" spins. This pattern greatly facilitates the detection of entanglement and its volume law distribution after dynamics. Entanglement, however, is a notoriously difficult entity to detect even for single pairs and require local measurements in at least two non-commuting bases. Thus we design the second chip which directly interferes distant output modes of the first chip to make entanglement detection  feasible through photon counting. This second chip will have a versatile application as a detector of entanglement
from the outputs of any other future chips that implement other (generic) spin chain dynamics. 

\section*{Theory}

Through our quantum transport chip, we simulate the dynamics induced by a ``quantum quench" -- a process in which the interactions inside a many-body system are suddenly changed. Let us consider preparing initially the famous N\'{e}el state $\ket{\psi_{\mathrm{Neel}}}=\ket{\downarrow \uparrow \downarrow  ... \uparrow \downarrow}$ of simulated spins, which is the ground state of the antiferromagnetic Ising chain and manifestly has no entanglement between any pairs of spins (the whole state is a separable state). In a condensed matter scenario, this state would be prepared by cooling under a Ising Hamiltonian. In photonic technology, we simply inject our apparatus with a simulated N\'{e}el state as to be described later. Immediately following the injection, the state starts evolving in accordance to the Hamiltonian simulated by our chip. This corresponds to the archetypical (perhaps the most extensively theoretically studied) spin chain quench \cite{Wichterich09,Banchi-Bose,DeChiara,DemlerQuench} in which the system Hamiltonian is suddenly changed from the Ising model to the XY model
\begin{equation}
H= \sum_{i=1}^{N-1} \frac{1}{2}J_i (X_i X_{i+1} + Y_i Y_{i+1}) 
\label{HXX}
\end{equation}
where $X_i$, $Y_i$ represent the Pauli matrices for the spin in the site $i$ and $N$ is the length of the chain. Further, the photonics platform enables us to set the $J_i=\sqrt{i(N-i)}$ (which has not yet been feasible in the atomic/ionic platforms for quench). Then one generates, after a relevant time $t^*$, a remarkable pattern of nested entangled states \cite{Banchi-Bose,Nested}, also dubbed as ``rainbow states" \cite{Rainbow} (cf Fig.\ref{concept}):
\begin{equation}
\ket{\psi_{\mathrm{v.l.}}}=\vert \psi^+_{1,N} \rangle \vert \psi^+_{2,N-1} \rangle...
\label{generic_target}
\end{equation}
where $\ket{\psi^+_{i,j}}=\frac{1}{\sqrt{2}}(\ket{\uparrow_{i} \downarrow_{j}}+\ket{\downarrow_{i} \uparrow_{j}})$. This state corresponds to a ``volume law" for entanglement as the number of entangled pairs is $\sim N/2$ (exactly $N/2$ for even $N$ and $(N-1)/2$ for the odd $N$). In other words, the amount of entanglement between the left and the right halves of the spin chain generated due to the quench dynamics scales as the size $\sim N/2$ of each part.  Even more strikingly, it corresponds to the maximal entropy of entanglement, a much discussed quantity in recent theory and experiment \cite{Islam15}, between the left and the right blocks of the spin chain. Remarkably, at time $2t^{*}$ the system implements a perfect QST \cite{Christandl04} and thus returns to its original separable state.

\begin{figure*}[ht!]
    \includegraphics[width=1\textwidth]{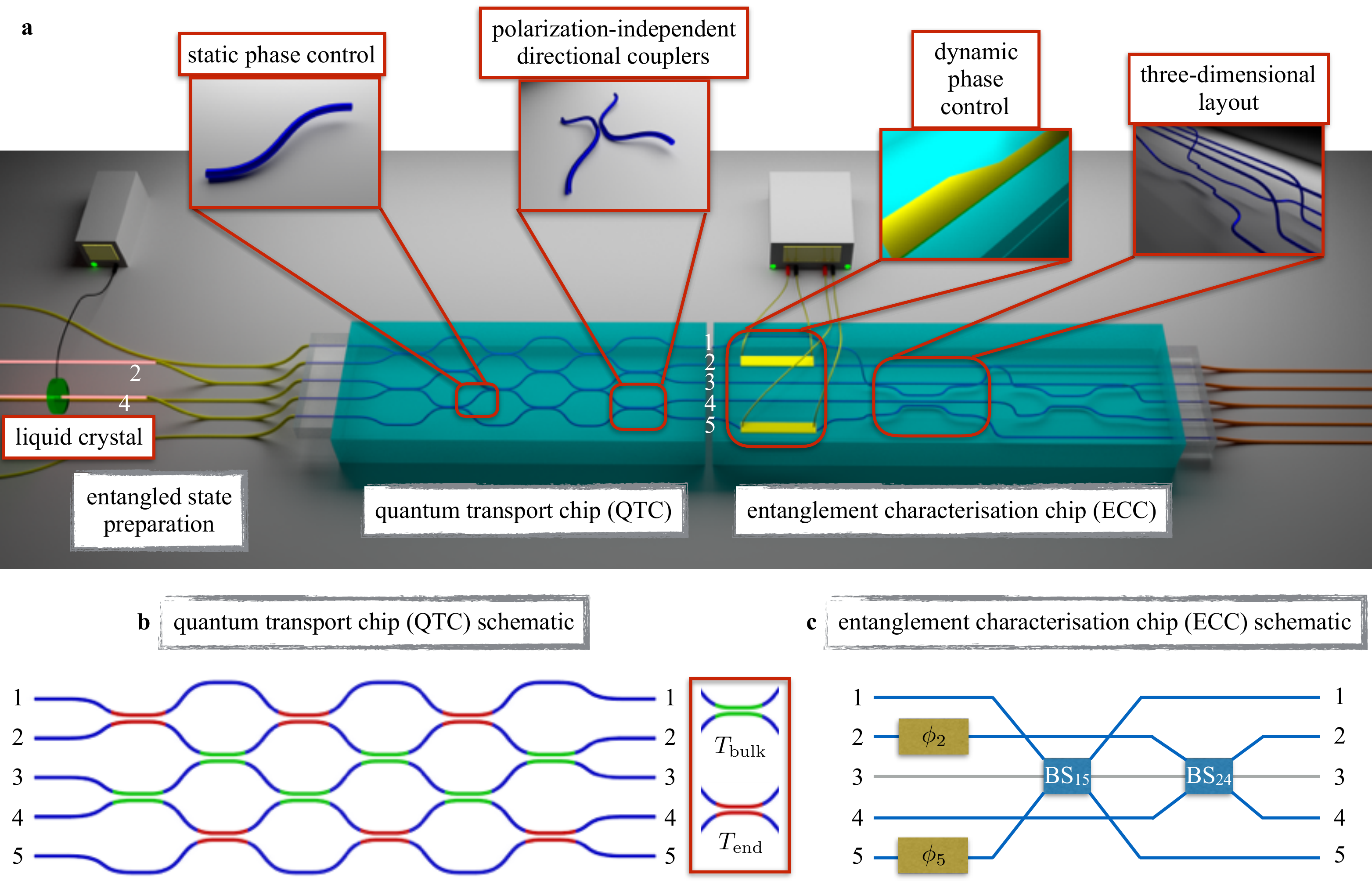}
 \caption{\textbf{a.} Representation of the experimental apparatus comprised of the first quantum transport chip and the second entanglement characterisation chip. The insets highlight the specific elements and the three dimensional geometry of the photonic devices. \textbf{b.} Schematisation of the quantum transport device of the first chip showing in green the directional couplers of the bulk (with transmissivity \textit{T}$_{\mathrm{bulk}}=0.36$) and in red those at the edge (with transmissivity \textit{T}$_{\mathrm{end}}=0.25$) of the device. \textbf{c.} Schematic representation of the entanglement characterisation device of the second chip, showing the dynamic phase controls $\phi_{2}$ and $\phi_{5}$ acting on the $\mathrm{2}^{\mathrm{nd}}$ and $\mathrm{5}^{\mathrm{th}}$ waveguides and the 50/50 beam splitters between the (2,4) and (1,5) pairs of modes. The 3$^{\mathrm{rd}}$ waveguide (in grey) is not involved in any interference processes.} \label{setup}
\end{figure*}

Our simulation technique combines three principal ingredients. Firstly, using
the Jordan-Wigner transformation \cite{sengupta,Calabrese-Cardy,Wichterich09,Banchi-Bose} it is possible to map spin excitations in a chain with nearest-neighbour interactions onto non-interacting fermions hopping in a lattice (hence the name quantum transport chip). Secondly, fermionic behaviour can be simulated in photonic platforms using two-photon entangled states with antisymmetric wavefunctions; this has already been exploited for simulating several phenomena, such as Bosonic-Fermionic continuous transition \cite{Sansoni12,Matthews13}, Anderson localisation \cite{CrespiAnderson} and Fano interference effect in quantum decay \cite{CrespiFano}. Third is the fact that while the spin chain dynamics maps effectively to a continuous time quantum walk of multiple fermions it can be simulated via discrete quantum walks in the weak transmission limit \cite{Strauch06}. Discrete quantum walks provide more accurate control over the implemented Hamiltonian \cite{Reck94} due to the additional degree of freedom given by the coin operator. An $N$-sites continuous quantum walk with site-to-site coupling $J_i$ can be simulated by a photonic network constituted by $M$ layers of beamsplitters in cascade (see Fig. \ref{concept}) with transmissivities $
T_i=\sin^2(\epsilon J_i)$, where the scaling factor $\epsilon=\frac{N+1}{M}=\frac{2 t^*}{M}$ should be small to be in the weak transmission limit, which means that the approximation will be more accurate with a larger number of steps. In essence, this third ingredient is thus a digital approach to quantum simulations as recently undertaken in superconducting networks \cite{Barends}.
Further details on the theoretical model are provided in the Supplementary Information.

\section*{Experiments}

In our work, the quantum transport chip is a 5-sites, 6-steps integrated interferometer with engineered couplings for approximating the dynamical generation of the rainbow state of Eq. (\ref{generic_target}) which displays the volume law.  We studied the evolution of a two-particle fermionic states (equivalent to the N\'{e}el state of a 5 site antiferromagnet) and its bosonic counterpart, analysing the different statistics at the output and confirming the prediction for creation of the rainbow pattern of entanglement in the fermionic case. This has been possible by merging several experimental techniques, such as entanglement generation in bulk optics, state propagation in polarisation-insensitive integrated circuits, and entanglement analysis in a 3D tuneable integrated circuit. 

\textit{Optical Circuits Fabrication}. 
The Quantum Transport Chip (QTC) consists of 5 waveguides that form a series of cascaded directional couplers (Fig. \ref{setup}a). Each waveguide represents a spin particle and the directional couplers operate as the integrated equivalent of beam splitters discretising the interactions between the spins. The presence of a photon in one waveguide encodes the spin state $\ket{\uparrow}$ while the absence of a photon represents $\ket{\downarrow}$.

In order to properly reproduce the discrete time quantum state transfer it is required to use two different transmission values, \textit{T}$_{\mathrm{bulk}}$ = 0.36 and \textit{T}$_{\mathrm{end}}$ = 0.25, for the internal and the external beam splitters respectively (see Fig. \ref{setup}b and Supplementary Information). This can be achieved by increasing the separation between the two waveguides in the interaction region of the external couplers.

The spin chain will be initiated using two photons prepared in a polarisation entangled state (see below). The chip needs to be polarisation insensitive so that the polarisation encoding does not directly affect the propagation of the two-photon state. By fabricating the couplers in a tilted geometry the transmissivity is equal for both horizontal and vertical polarisation \cite{Sansoni12}.

The Entanglement Characterisation Chip (ECC) is a device designed to collect and interfere the outgoing signals from the QTC, allowing the evaluation of the entanglement that is generated in the latter (as seen in Fig. \ref{setup}a).
It consists of two polarisation insensitive directional couplers formed by waveguides 1-5 \& 2-4 that collect and interfere the corresponding outputs from the first device, and two heating elements fabricated over waveguides 2 and 5 in order to scan the interference fringes by modulating the phase (Fig. \ref{setup}c) \cite{Flamini01, Meany01}. 

The ECC adopts a fully three-dimensional design. The waveguides start at a shallow depth of 25 $\mu$m on the collection side so the heating elements will operate efficiently. Then they go deeper at 65 $\mu$m to form the couplers without any physical crossing of the waveguides (see Fig. \ref{setup}a inset). The central waveguide is not interacting with any other one.

\textit{Measurements and Results}. The simulation of the spin chain dynamics on the photonic platform has been performed by exploiting polarisation-entangled pairs of photons, generated by type-II parametric downconversion process in a BBO (beta barium borate) crystal. The general expression for the state of the generated photons can be written as $\ket{\Psi^{\chi}_{2 \mathrm{p}}} = 2^{-1/2} (\ket{HV} + e^{\imath\chi}\ket{VH})$, 
where the subscript $2 \mathrm{p}$ indicates a two-photon state. The phase parameter $\chi$ can be controlled by means of an electrically configurable liquid crystal positioned on one of the two photon paths. Thus both bosonic and fermionic statistics can be simulated by using the Bell states $\ket{\Psi^{+}_{2 \mathrm{p}}}$ and $\ket{\Psi^{-}_{2 \mathrm{p}}}$ respectively for their symmetric and anti-symmetric wavefunctions \cite{Bose06,Sansoni12}. 

\begin{figure}[ht!]
    \includegraphics[width=0.5\textwidth]{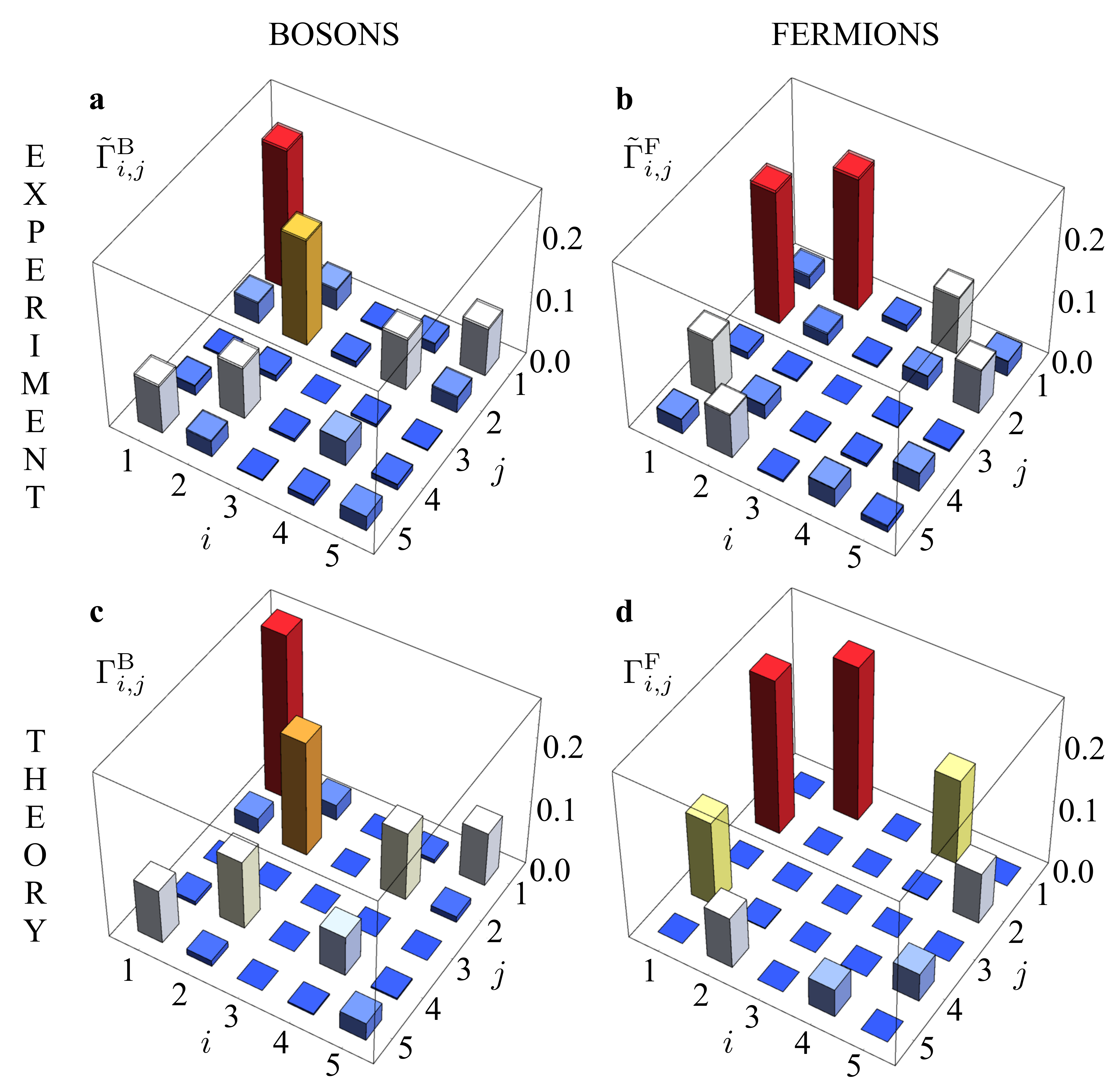}
 \caption{\textbf{a-b.} Experimental results of the correlation function, $\tilde{\Gamma}_{ij}$, for bosonic and fermionic transport. \textbf{c-d.} Theoretical prediction for the correlation function, for bosonic and fermionic transport obtained from the unitary matrix \textit{U} of the transport device.} \label{BosVsFer}
\end{figure}

\begin{figure}[hb!]
\includegraphics[width=0.5\textwidth]{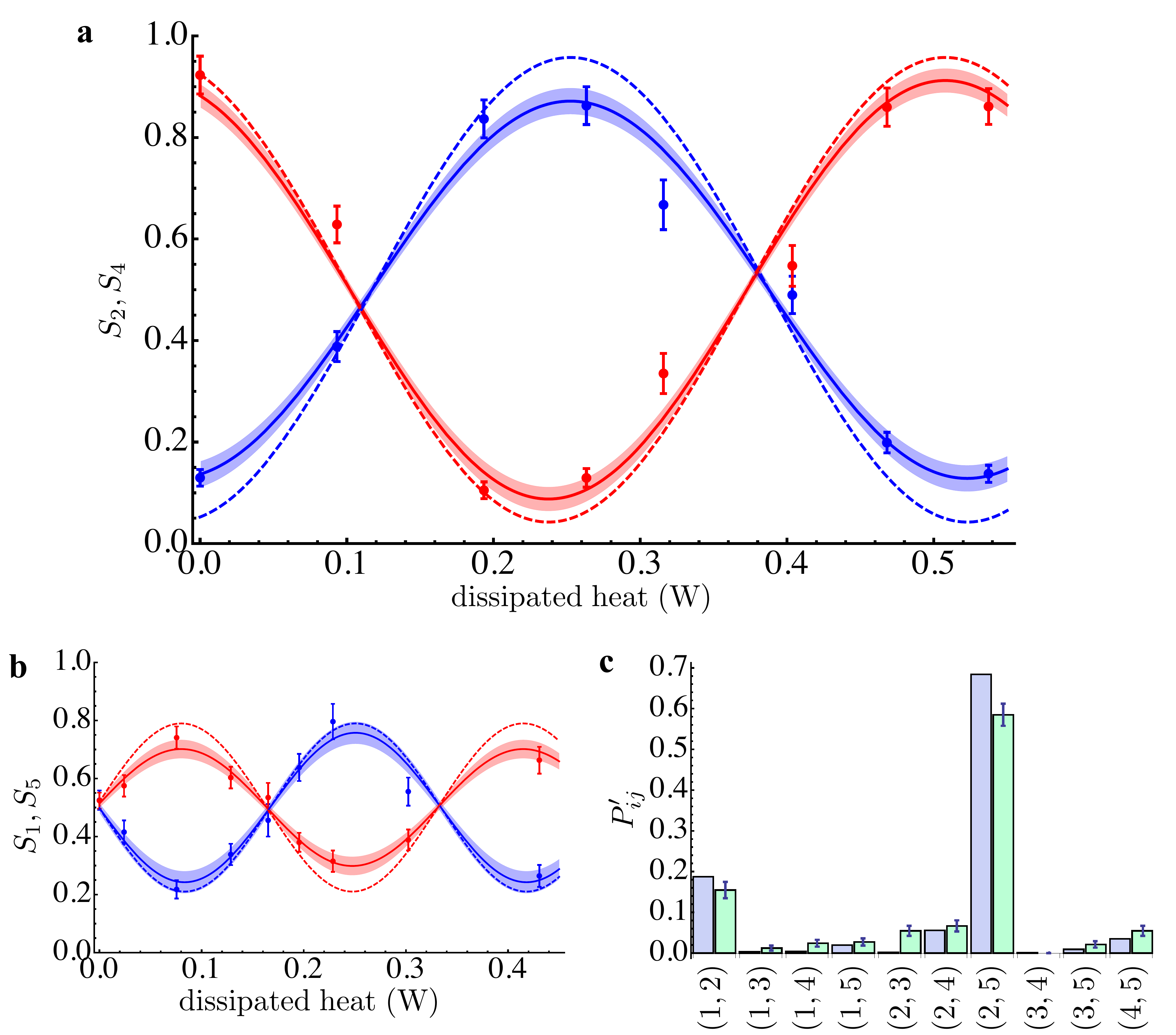}
\caption{Interference fringes for $S_2$ (blue points) and $S_4$ (red points) as a function of the dissipated heat on mode 2 (proportional to $\phi_{2}$). \textbf{b.} Interference fringes for $S_1$ (blue points) and $S_5$ (red points) as a function of the dissipated heat on mode 5 (proportional to $\phi_{5}$). In both sub-panels \textbf{a} and \textbf{b}, solid lines and shaded areas represent respectively the best fit curves and fit curve error, while the dashed lines corresponds to theoretical predictions. \textbf{c.} Green bars: two-photon probability distribution obtained at the output of the second chip, when the polarisation-entangled anti-symmetric state is injected into input 2 and 4 of the first chip. The distribution is obtained for values of the two phases which maximise $P'_{25}$; blue bars: theoretical predictions. 
All theoretical models are obtained taking into account the reconstructed matrices of the first chip ($\tilde{U}_{H}$ and $\tilde{U}_{V}$) and the coupling efficiencies at the interface between the two devices (see Supplementary Information).}
 \label{entanglement}
\end{figure}

 The N\'{e}el state $\ket{\downarrow \uparrow \downarrow \uparrow \downarrow}$ of the spin chain was simulated by injecting the polarisation-entangled state $\ket{\Psi^{-}_{2 \mathrm{p}}}$ in the inputs 2 and 4 of the QTC (see Fig. \ref{setup}a-b) through a single mode fiber array and, after evolution into the interferometers, photons are collected via a multimode fiber array and sent to single photon avalanche photodiodes (APDs). At the output we collect coincidence measurements from all pairs of outputs to measure the off-diagonal elements (outputs $i\neq j$); for the diagonal contributions (outputs $i=j$) an in-fiber beam-splitter is inserted in each single output mode. The same experiment is also performed with $\ket{\Psi^{+}_{2 \mathrm{p}}}$ to capture the distinctive features of bosonic and fermionic dynamics on the QTC.

In the limit of a QTC with an infinite number of steps, one would expect a perfect generation of the entangled state of the form (\ref{generic_target}) exhibiting the volume law at $t^*$. A device with 6 steps (whose unitary we denote $U$), when injected with a N\'{e}el state, would give unbalanced $\ket{\psi^{+}}$ states at symmetric sites (see below). Finally, one has to take into account imperfections of the implemented unitary, by considering also the possible slight difference when acting on photons with different polarisations ($\tilde{U}_{H}$ and $\tilde{U}_{V}$).

As a first step, tomographic reconstruction of the unitary transformation matrix has been performed for both polarisations \textit{H} and \textit{V}, which gave fidelities with respect to the expected unitary of $\mathcal{F}^{H} = 0.962 \pm 0.001$ and $\mathcal{F}^{V} = 0.977 \pm 0.002$, proving the high quality and the polarisation independence of the device (see Supplementary Information for more details on the unitary characterisation).

The results of the quantum dynamics are shown in Fig. \ref{BosVsFer} for both bosonic and fermionic behavior, where the experimental ($\tilde{\Gamma}^{\mathrm{B}/\mathrm{F}}_{ij}$) and theoretical ($\Gamma^{\mathrm{B}/\mathrm{F}}_{ij}$, calculated from the 6-steps unitary $U$) correlation functions \cite{Bromberg09,Peruzzo10} are reported. The output probabilities detected experimentally are related to the correlation function by the equations $\tilde{\Gamma}_{ii} = P_{ii}$ and $\tilde{\Gamma}_{ij} = \frac{P_{ij}}{2}$, due to the symmetrisation of the correlation function matrix. 

From the results shown in Fig. \ref{BosVsFer} one can observe the different bosonic and fermionic evolution: for states which are symmetric with respect to particle exchange, the bosonic distribution has a strong contribution on the diagonal terms of the matrix, which is a clear sign of the bosonic coalescence; for the anti-symmetric state, the resulting fermionic evolution gives strong contributions on the off-diagonal terms of the matrix. 

Each experimental distribution is compared to its theoretical prediction through the measurement of the similarity, with results for the bosonic and fermionic transport of $\mathcal{S}^{\mathrm{B}} = 0.942\pm0.005$ and $\mathcal{S}^{\mathrm{F}} = 0.836\pm0.004$. 

\textit{Entanglement Certification}. The time evolution of the state $\ket{\Psi^{-}_{2 \mathrm{p}}}_{24}$ (which simulates the N\'{e}el state) through an effective Hamiltonian of the form (\ref{HXX}) acting on simulated spins gives a state equal to $\ket{\psi^{+}_{1 \mathrm{p}}}_{15}\ket{\psi^{+}_{1 \mathrm{p}}}_{24}\ket{0}_3$ at time $t^{*}$, where $\ket{\psi^{+}_{1 \mathrm{p}}}_{ij}$ denotes the one-photon path-encoded Bell state $\ket{\psi^{+}}$ over the modes $i$ and $j$ \cite{Lombardi02}. This state simulates the volume law entangled state $\ket{\psi_{\mathrm{v.l.}}}$ of Eq.(\ref{generic_target}). As we perform the evolution with a finite number of steps, we get a state of the form $
\ket{\psi_{\mathrm{out}}} = (\alpha\ket{10}_{15}+\beta\ket{01}_{15}) (\gamma\ket{10}_{24}+\delta\ket{01}_{24})\ket{0}_3$. The imbalance between $\alpha$ and $\beta$ and between $\gamma$ and $\delta$, which can be observed in Fig. \ref{BosVsFer}, comes from the fact that a finite number of discrete step is an approximation of a perfect state transfer. To check the coherence between the terms $\ket{10}_{ij}$ and $\ket{01}_{ij}$, a phase shift and a beamsplitter transformation over the modes $i$ and $j$ can be applied, where $(i,j)=(1,5)$ and $(2,4)$. This transformation is implemented in the ECC device (see Fig. \ref{setup}c). 

We first measured the two-photon interference fringes obtained by varying the phases $\phi_2$ and $\phi_5$. More specifically, we defined the quantities $S_{i}$ (see Methods) which depend only from the tuning of a single phase, while being insensitive to possible cross-talks over neighbouring modes due to heat dispersion. Interference fringes, shown in Fig. \ref{entanglement}a-b, are clearly visible when plotting $S_1$ and $S_5$ ($S_2$ and $S_4$) as a function of $\phi_5$ ($\phi_2$).

To further characterise the degree of entanglement, we evaluated the {\it
entanglement fraction} of systems 1-5 and 2-4 with respect to the ideal
one-photon Bell state $\ket{\psi^{+}_{1 \mathrm{p}}}$: $\mathcal E_{ij}=\, _{ij}
\langle \psi^{+}_{1 \mathrm{p}} \vert \, \rho_{ij} \, \vert \psi^{+}_{1
    \mathrm{p}} \rangle_{ij}$, where $\rho_{ij}$ is the reduced density matrix
    for two qubits in positions $i$ and $j$ \cite{Wootters96}. These quantities
    can be evaluated by measuring the single-photon probabilities $N'_i$ and
    two-photon probabilities $P'_{ij}$ after evolution through the second
    device. Indeed, one can find that $\mathcal E_{15}=N'_5-P'_{15}$ and
    $\mathcal E_{24}=N'_2-P'_{24}-P'_{23}+P'_{34}$ (See Supplementary
    Information). The first expression implies finding a photon in mode 5 and no
    photons in mode 1. The second one keeps nonlocal terms from the mapping from
    spin excitations to fermions, which in the case of $\mathcal E_{15}$ are
    factorised into a fixed phase term.  By evaluating the entanglement
    fractions from the two-photon probabilities shown in Fig.
    \ref{entanglement}c we obtain $\mathcal E_{15} = 0.66 \pm0.03$ and
    $\mathcal E_{24} = 0.74 \pm 0.03.$ This amounts to an approximate  verification of Eq.(\ref{generic_target}) and thereby the growth of entanglement to a volume law.

\section*{Conclusions and Discussion}  

Using integrated photonics we have achieved the first quantum simulation of the extensive growth of entanglement in a spin chain after a quench. The volume law for the amount of entanglement between the two halves of the spin chain after an appropriate time $t^{*}$ is verified. 
While atomic simulators have the verification of the above as one of the chief
items in their future roadmap \cite{Islam15}, the exquisite control of couplings
offered by integrated photonics has made it already feasible. Moreover, our
methodology of designing the QTC, in particular, a time discretisation of the
dynamics that we exploit, makes it amenable to alter the chip design to simulate
a wide range of such spin chain dynamics -- e.g., time dependent quantum control
\cite{Rabitz}, and, if some nonlinearities can be potentially introduced to the
waveguides \cite{Nonlinear}, then a much wider range of spin Hamiltonians such
as isotropic models. The ECC, on the other hand, is versatile -- it will enable a detection of the entangled outputs from any of the above range of spin chain dynamics that can be accomplished in integrated photonics. In the future, scaling to larger spin chains should also be possible with fully antisymmetric states of more photons \cite{Matthews13}, for example using additional degrees of freedom, mimicking more fermions. The entanglement growth we have observed is responsible for thermalisation in systems that do thermalise, and also corresponds to the formation of a black hole through the holographic correspondence. Aside this fundamental interest, the level of control shown by the generation of the rainbow states with a high entangled fraction, will make integrated photonics an ideal platform to benchmark and optimise quantum nano-devices of the future.

\section*{Methods}

\textit{Devices fabrication technique.} In both chips, the QTC and the ECC, the waveguides are inscribed by femtosecond laser writing in Corning Eagle-2000 borosilicate glass. We used a femtosecond Yb:KYW cavity-dumped oscillator at 1030 nm wavelength, emitting pulses with 300 fs duration, and at a 1 MHz repetition rate.  The laser beam was focused through a 50x, 0.6 NA microscope objective lens into the glass substrate, which was translated by a computer-controlled three-axis Aerotech FiberGlide 3D series stage, at a velocity of 40 mm/s. In the QTC the waveguides are inscribed using 220 mW of laser power at a depth of 170 $\mu$m. In the ECC the power of inscription was slightly lower at 210 mW since the waveguides were closer to the surface.

\textit{QTC characterisation.} After fabrication, the circuit was characterised classically using a diode laser. The overall performance was verified by collecting the output power distribution for each individual input and comparing it to the theoretically expected one. The polarisation response of the chip was identical for both vertical and horizontal polarisations within 2\%, and the similarity between the theoretically expected and experimentally measured output distributions were $\mathcal{S}^{H} = 0.963 \pm 0.002$ and $\mathcal{S}^{V} = 0.976 \pm 0.002$.

\textit{ECC characterisation.} Before using the ECC in the actual quantum experiment, the device was characterised independently by coupling it to a chip containing balanced directional couplers, thus forming Mach-Zehnder interferometers. By applying different voltages to the heating electrodes it was possible to observe the interference fringes at the output of the corresponding interferometers with visibilities $V_{1-5} = 0.86 \pm 0.02$ and $V_{2-4} = 0.94 \pm 0.02$ (for an expect value of $V_{i-j} = 1$). In addition, we verified the stability of the thermal tuning of the phase and we have observed that a constant phase value could be maintained for a period of 8 hours with a standard deviation $\sigma_{\Phi} = |0.03|$ radians.

\textit{Interference fringes measurement.} Let us define $S_{1} = P'_{12}+P'_{14}$, where $P'_{ij}$ is the two-photon probability at the outputs $i$ and $j$ of the second device. Such quantity depends only from the tuning of phase $\phi_{5}$ on mode $5$, while being insensitive to possible thermal cross-talks between the modes. Analogous quantities can be defined for modes $2$, $4$ and $5$ ($S_2= P'_{21}+P'_{25}$, $S_4= P'_{41}+P'_{45}$ and $S_5= P'_{52}+P'_{54}$). The recorded visibilities were $V_{S_{1}}=0.51\pm0.05$, $V_{S_{5}}=0.40\pm0.03$, $V_{S_{2}}=0.74\pm0.03$, and $V_{S_{4}}=0.82\pm0.03$.

\textbf{Acknowledgements.} 
This work was supported by the Marie Curie Initial Training Network PICQUE (Photonic Integrated Compound Quantum Encoding, grant agreement no. 608062, funding Program: FP7-PEOPLE-2013-ITN, http://www.picque.eu) and by the H2020-FETPROACT-2014 Grant QUCHIP (Quantum Simulation on a Photonic Chip; grant agreement no. 641039, http://www.quchip.eu).

\textbf{Author contribution.} L.B., S.B., P.M., R.O., F.S. conceived the idea. I.P., A.C., R.O. designed and fabricated the integrated photonics devices and carried out their characterization with classical light. A.S.R., D.C., M.B., N.S., F.S. carried out the quantum experiments and analyzed the data. L.B, S.B. developed the theoretical model of engineered quantum walk and of entanglement growth. All authors discussed the experimental implementation and results, and contributed to writing the paper.

\textbf{Competing financial interests.}
The authors declare no competing financial interests.

\textbf{Materials \& Correspondence.} 
Correspondence and requests for materials should be addressed to S.B., R.O and F.S.

\end{document}


\title{Photonic Simulation of Entanglement Growth After a Spin Chain Quench\\
- Supplementary Information-}

\author{Ioannis Pitsios}
\affiliation{Istituto di Fotonica e Nanotecnologie - Consiglio Nazionale delle Ricerche (IFN-CNR), P.za Leonardo da Vinci, 32, I-20133 Milano, Italy}
\affiliation{Dipartimento di Fisica - Politecnico di Milano, P.za Leonardo da Vinci, 32, I-20133 Milano, Italy}

\author{Leonardo Banchi}
\affiliation{Department of Physics and Astronomy, University College London, Gower Street, WC1E 6BT London, United Kingdom}

\author{Adil S. Rab}
\affiliation{Dipartimento di Fisica - Sapienza Universit\`{a} di Roma, P.le Aldo Moro 5, I-00185 Roma, Italy}

\author{Marco Bentivegna}
\affiliation{Dipartimento di Fisica - Sapienza Universit\`{a} di Roma, P.le Aldo Moro 5, I-00185 Roma, Italy}

\author{Debora Caprara}
\affiliation{Dipartimento di Fisica - Sapienza Universit\`{a} di Roma, P.le Aldo Moro 5, I-00185 Roma, Italy}

\author{Andrea Crespi}
\affiliation{Istituto di Fotonica e Nanotecnologie - Consiglio Nazionale delle Ricerche (IFN-CNR), P.za Leonardo da Vinci, 32, I-20133 Milano, Italy}
\affiliation{Dipartimento di Fisica - Politecnico di Milano, P.za Leonardo da Vinci, 32, I-20133 Milano, Italy}

\author{Nicol\`o Spagnolo}
\affiliation{Dipartimento di Fisica - Sapienza Universit\`{a} di Roma, P.le Aldo Moro 5, I-00185 Roma, Italy}

\author{Sougato Bose}
\affiliation{Department of Physics and Astronomy, University College London, Gower Street, WC1E 6BT London, United Kingdom}

\author{Paolo Mataloni}
\affiliation{Dipartimento di Fisica - Sapienza Universit\`{a} di Roma, P.le Aldo Moro 5, I-00185 Roma, Italy}

\author{Roberto Osellame}
\affiliation{Istituto di Fotonica e Nanotecnologie - Consiglio Nazionale delle Ricerche (IFN-CNR), P.za Leonardo da Vinci, 32, I-20133 Milano, Italy}
\affiliation{Dipartimento di Fisica - Politecnico di Milano, P.za Leonardo da Vinci, 32, I-20133 Milano, Italy}

\author{Fabio Sciarrino}
\affiliation{Dipartimento di Fisica - Sapienza Universit\`{a} di Roma, P.le Aldo Moro 5, I-00185 Roma, Italy}

\maketitle

\onecolumngrid


\def\sx{6}
\def\sy{5}

\newcommand{\bs}[2]{
  \pgfmathtruncatemacro{\X}{#1}
  \pgfmathtruncatemacro{\Y}{#2}
  \begin{scope}[shift={(\sx*\X,\sy*\Y)}]
    \draw [blue, line width=10pt] plot [smooth,tension=0.5] coordinates 
    {(1,0) (1+.1,1.3) (2,2) (2,3) (1+.1,3.7) (1,5)};
    \draw [blue, line width=10pt] plot [smooth,tension=0.5] coordinates 
    {(4,0) (4-.1,1.3) (3,2) (3,3) (4-.1,3.7) (4,5)};
    \node at (2.5,0) {\huge output};
    \node at (2.5,5) {\huge input};
  \end{scope}
}

\newcommand{\inposition}[3]{
  \pgfmathtruncatemacro{\X}{#1}
  \pgfmathtruncatemacro{\Y}{#2}
  \begin{scope}[shift={(\sx*\X,\sy*\Y)}]
    #3
  \end{scope}
}

\newcommand\drawLL[1]{
  \draw [#1, line width=10pt] plot [smooth,tension=0.5] coordinates 
  {(1,0) (1+.1,1.3) (2,2) (2,3) (1+.1,3.7) (1,5.1)};
}
\newcommand\drawLR[1]{
  \draw [#1, line width=10pt] plot [smooth,tension=0.5] coordinates 
  {(4,0) (4-.1,1.3) (3,2) (3,3) (4-.1,3.7) (4,5.1)};
}
\newcommand\drawRL[1]{
  \draw [#1, line width=10pt] plot [smooth,tension=0.5] coordinates 
  {(1,0) (1-.1,1.3) (0,2) (0,3) (1-.1,3.7) (1,5.1)};
}
\newcommand\drawRR[1]{
  \draw [#1, line width=10pt] plot [smooth,tension=0.5] coordinates 
  {(4,0) (4+.1,1.3) (5,2) (5,3) (4+.1,3.7) (4,5.1)};
}

\newcommand\Lbound{
  \filldraw[fill=white,color=white] 
  (-2.5, 0) -- (1.2,0) -- (2.5,2.5) -- (1.2,5.1) -- (-2.5,5.1);
  \draw[ultra thick, dashed] 
  (1.2,0) -- (2.5,2.5) -- (1.2,5);
}
\newcommand\Rbound{
  \filldraw[fill=white,color=white] 
  (6.5, 0) -- (3.8,0) -- (2.5,2.5) -- (3.8,5.1) -- (6.5,5.1);
  \draw[ultra thick, dashed] 
  (3.8,0) -- (2.5,2.5) -- (3.8,5);
}

\newcommand{\bsL}[5]{
  \pgfmathtruncatemacro{\V}{#3}
  \inposition{#1}{#2}{
    \draw [fill=orange!50,orange!50] (0,1.5) rectangle (6,3.5);
    \drawLL{#4} \drawLR{#5}
    \node at (2.5,4) { \huge\bf \V };
  }
}
\newcommand{\bsR}[5]{
  \pgfmathtruncatemacro{\V}{#3}
  \pgfmathtruncatemacro{\Z}{#3+2}
  \inposition{#1}{#2}{
    \draw [fill=orange!50,orange!50] (0,1.5) rectangle (6,3.5);
    \drawRL{#4} \drawRR{#5}
    \node at (-0.5,4) { \huge\bf \V };
    \node at (5.5,4) { \huge\bf \Z };
  }
}

\newcommand{\bsLm}[5]{
  \inposition{#1}{#2}{
    \draw [fill=orange!50,orange!50] (0,1.5) rectangle (6,3.5);
    \drawLL{#4} \drawLR{#5}
    \node at (2.5,4) { \huge\bf #3 };
  }
}
\newcommand{\bsRm}[5]{
  \inposition{#1}{#2}{
    \draw [fill=orange!50,orange!50] (0,1.5) rectangle (6,3.5);
    \drawRL{#4} \drawRR{#5}
    \node at (-0.5,4) { \huge\bf #3 };
  }
}

\section*{Discretization of continuous time quantum walks}
\begin{suppFig}[ht]
  \begin{center}
  \scalebox{0.3}{
    \begin{tikzpicture}
      \foreach \x in {0,...,4}{
	   \foreach \y in {0,...,1}{
	    \bsR{\x}{2*\y}{2*\x}{blue}{blue}
	    \bsL{\x}{2*\y+1}{2*\x+1}{blue}{blue}
	   };
      };
    \end{tikzpicture}
  }
  \end{center}
  \caption{
      Example configuration with $9$ sites and 4 time steps.  There are also two
      boundary sites 0,10.  The spatial
      direction corresponds to the horizontal one while the ``time'' is the
      vertical direction. Light come from top to bottom.  
      Each beam splitter is localized in a { site} $i$.
  }
  \label{fig:BS}
\end{suppFig}

We consider a configuration as in Supplementary Fig.\ref{fig:BS} and we 
 call $\ket\la$ and $\ket\ra$ the states moving leftwards or rightwards. 
The left input of the beam splitter (BS) $x$ is named $\ket{x,\ra}$, while the right
input is named $\ket{x,\la}$. On the other hand, the right output 
of BS $x$ becomes the left input of BS $x+1$, namely
$\ket{x{+}1,\ra}$. Similarly, the left output of BS $x$ is 
$\ket{x{-}1,\la}$. In summary,
\begin{equation}
\begin{aligned}
  \ket{{\rm in},x,\LA} &= \ket{x,\ra}, &
  \ket{{\rm in},x,\RA} &= \ket{x,\la}, \\  
  \ket{{\rm out},x,\LA} &= \ket{x{-}1,\la}, &
  \ket{{\rm out},x,\RA} &= \ket{x{+}1,\ra}.
  \label{e.mapping}
\end{aligned}
\end{equation}
Each beam splitter implements the transformation $B$, defined as 
$
  B|{\RA}\rangle = t'|{\LA}\rangle + r'|{\RA}\rangle,
  B|{\LA}\rangle = r |{\LA}\rangle + t |{\RA}\rangle, 
$
so in general
\begin{equation} 
\begin{aligned}
  B_x\,|{x, \la}\rangle &= t'_x\,|{x{-}1,\la}\rangle + r'_x\,
  |{x{+}1,\ra}\rangle~, &
  B_x\,|{x,\ra}\rangle &= r_x\, |{x{-}1,\la}\rangle + t_x\, |{x{+}1,\ra}\rangle 
  ~.
  \label{e.bs}
\end{aligned}
\end{equation}
We define the evolution operator $U=\oplus_x B_x$ generated by the series of
beam splitters. This operator can be split
in two terms, $U=SC$, 
as in the quantum walk literature, where
$S=\sum_x \ket{x{-}1}\bra{x} \otimes\ket\la\bra\la 
\,+\,\ket{x{+}1}\bra{x} \otimes\ket\ra\bra\ra$ is the shift operator, and $C$ is a
{\it coin operator}
$
  C\ket{x,\la} = t'_x \ket{x,\la} + r_x \ket{x,\ra}, 
  C\ket{x,\ra} = r'_x \ket{x,\la} + t_x \ket{x,\ra}, 
$
obtained by reshuffling the beam splitter operator. By parametrizing each beam
splitter operator with its transmissivity $T_x=t_x^2$ and by defining the angle 
$\Theta_x$ such that  $T_x=\sin^2\Theta_x$, one can show that the coin operator 
can be written as 
$
  C = \exp\left[i \left(\frac\pi2\openone -
  \hat\Theta\right)\,\sigma^y\right],
$
where
 $\hat{\Theta} =
\sum_x \hat\Theta_x \ket x \bra x$ and where $\sigma^y$ is the Pauli matrix
acting on the coin space.

To simplify the theoretical analysis we consider periodic boundary conditions.
This does not affect the generality of the result and, for instance, 
open boundary conditions can be obtained by adding an auxiliary BS with zero transmittance, 
$T_{\rm Aux}=0$. Then it is possible to define the Fourier basis 
$
  \ket{\tilde p} = \frac1{\sqrt{N+1}}\sum_x e^{-i \tilde p x}
  \ket{x},~ \tilde p = 2\pi\frac{p}{N+1}~ 
  \label{e.fourier}
$
which diagonalizes the shift operator. Indeed, in terms of the global 
momentum operator
$\hat P=\sum \tilde p\ket{\tilde p}\bra{\tilde p}$, it is
$
S=\exp\left(-i \hat P \sigma^z\right)~.
$
As interference effects can occur only 
between even/odd sites, it is convenient to study $U^2$ and 
since $e^{i \pi/2 \sigma^y} e^{-i \hat P \sigma^z} e^{i \pi/2 \sigma^y}
= -  e^{i \hat P \sigma^z}$ and $e^{-i \hat\Theta\sigma^y}
= e^{-i \pi/2 \sigma^z} e^{i \hat \Theta \sigma^y} e^{i \pi/2 \sigma^z}$ 
one obtains
\begin{equation}
  U^2 = -
  \exp\left(-i \hat P \sigma^z\right)
  \exp{\left[-i \hat\Theta\sigma^y\right]}
  \exp{\left[i \left(\hat P - \frac\pi2\openone\right)\,\sigma^z\right]}
  \exp{\left[i \hat\Theta\sigma^y\right]}
  \exp{\left[i \frac\pi2\openone\sigma^z\right]}~.
  \label{e.U}
\end{equation}
The above  exact equation is the starting point for the theoretical
approximations leading to the simulation of continuous time quantum walks.

When $T_x = 1-\epsilon_x$, with $\epsilon_x \ll 1$, the evolution 
generated by Eq.\eqref{e.U} in the long-time and long-distance
limit can be approximated by the evolution generated by a Dirac 
Hamiltonian 
\cite{strauch2006connecting,kurzynski2011discrete}.
This limit is somehow trivial for the purpose of quantum state transfer.
Photons basically move only in one direction, since the reflectivity
is negligible, and there is no complex interference phenomena.

The interesting case arises in the opposite limit $T_x\ll 1$,
namely when the beam splitters have a {low transmittance}. 
In this limit, it has been shown in 
\cite{strauch2006connecting,kurzynski2011discrete} 
that one can simulate a continuous time quantum walk (CTQW) with a 
discrete time quantum walk (DTQW). 
However, those mappings require both $2N$ effective sites 
for simulating a CTQW with $N$ sites and also require a mixture of two types of beam
splitters, one type with $T_x\approx 1$ and another set with 
$T_x \approx 0$. Here we use the exact Eq.\eqref{e.U} to derive a 
{simpler mapping}. Using a first order expansion for small $\Theta_x$ one finds
\begin{equation}
\begin{aligned}
  U^2 &\simeq 
   - W\,\exp\left(i 2 \,\hat H\, \sigma^z\right)\,W^\dagger ,
   &   W &= \frac1{\sqrt2}\begin{pmatrix}
    i e^{-i\hat P} & -i e^{-i\hat P} \\ \openone&\openone
  \end{pmatrix},
\end{aligned}
\end{equation}
where we have defined the Hamiltonian as $2 H= e^{i\hat P} \hat\Theta +
\hat\Theta e^{-i\hat P} + \mathcal O(\hat\Theta^2)$. 
If $\Theta_x = \epsilon j_x$ where $\epsilon \ll 1$ then
\begin{equation}
\begin{aligned}
   H \approx \frac\epsilon2 \sum_x j_x
  \,\ket{x}\bra{x{+}1} + \text{h.c.}
  && j_x \approx \frac{\Theta_x}\epsilon
  \approx \frac{\sqrt{T_x}}\epsilon~.
  \label{e.Hamiltonian}
\end{aligned}
\end{equation}
Therefore, the effective evolution reads 
\begin{equation}
  U(t) \simeq U^t 
    = i^t W\,e^{it \,\hat H\, \sigma^z}\,W^\dagger ~.
\end{equation}
showing that a CTQW can be approximated with a discrete time one.


\section*{Discrete-time Quantum Transport} 

The dynamics of 
{ fully engineered chains} 
\cite{christandl2004perfect}, 
where
$
      j_x = \frac{\pi}{N+1} \sqrt{x(N-x)}~,
      x=1,\dots,N-1
$
and $N$ is the length of the chain, generates a perfect mirror of the 
initial state at the transmission 
time $2t^* = N+1$. Namely, every ``walker'' initially 
    in position $x$ is perfectly transferred to the position $N-x+1$, 
    after a time $t=2t^*$.
On the other hand, {minimally engineered chains} \cite{banchi2011long} require only 
the engineering of the transmissivity at the boundaries 
$ j_1 = j_{N-1}=j_{\rm opt}, j_x = 1$ for $ x=2,\dots,N-2 $, but allow only a 
high-quality (almost perfect) transmission between the two ends. 
For large $N$ there is an approximated formula 
\cite{banchi2011long} for the optimal coupling and the transmission
time: $j_{\rm opt} = 1.030 \,N^{-1/6}$, while $2t^*=N+1+2.29 \,N^{1/3}$.

\begin{suppFig}[ht]
\begin{center}
  \scalebox{0.3}{
    \begin{tikzpicture}
        \foreach \y in {0,...,1}{
          \bsLm{0}{2*\y+1}{1}{black}{blue}
          \bsRm{0}{2*\y}{Aux}{black}{blue}
          \bsL{1}{2*\y+1}{3}{blue}{blue}
          \bsR{1}{2*\y}{2}{blue}{blue}
          \bsL{2}{2*\y+1}{5}{blue}{blue}
          \bsR{2}{2*\y}{4}{blue}{blue}

          \bsRm{4}{2*\y+1}{N-6}{blue}{blue}
          \bsLm{4}{2*\y}{N-5}{blue}{blue}
          \bsRm{5}{2*\y+1}{N-4}{blue}{blue}
          \bsLm{5}{2*\y}{N-3}{blue}{blue}
          \bsRm{6}{2*\y+1}{N-2}{blue}{black}
          \bsLm{6}{2*\y}{N-1}{blue}{black}
          \inposition{6}{2*\y+1}{\node at (5.5,4) { \huge\bf Aux };}
        };

        \foreach \y in {0,...,3}{
            \inposition{2}{\y}{\Rbound}
            \inposition{4}{\y}{\Lbound}
        }

        \inposition{0}{4}{
          \node [single arrow, rotate=-90, fill=green!50] at (1,2.5) 
          {\bf \Huge ~ input ~};
          \node [single arrow, rotate=-90, fill=purple!50] at (4,2.5) 
          {\bf \Huge ~ input ~};
        }
        \inposition{6}{0}{
          \node [single arrow, rotate=-90, fill=green!50] at (4,-2.2) 
          {\bf \Huge ~ output ~};
          \node [single arrow, rotate=-90, fill=purple!50] at (1,-2.2) 
          {\bf \Huge ~ output ~};
        }

        \inposition{3}{4}{\node [scale=2] at (3,2.5) {\bf \Huge $N$ odd};}
        \inposition{3}{1}{
          \node [single arrow, scale=1.5, rotate=-90, fill=yellow!30] 
          at (3,5) {\bf \Huge even time steps $M$};}

    \end{tikzpicture}
  }
\end{center}
\caption{Simulation of perfect state transfer with $M$ time steps ($M$ odd). }
\label{fig:pst1}
\end{suppFig}

We first focus on perfect state transfer and we consider its simulation 
with $M$ discrete time steps. This requires to set $2t^*/\epsilon=M$, namely 
\begin{equation}
  \epsilon=\frac{2t^*}M=\frac{N+1}M~.
  \label{e.tsM}
\end{equation}
We consider an initial state in site $1$ which is an arbitrary superposition of 
the two ``coin''  degrees of freedom 
\begin{equation}
  \ket{\rm init} = \alpha \ket{1,\ra} + \beta \ket{1,\la}~.
  \label{e.init}
\end{equation}
When $M\gg N$, after $M$ steps one finds 
\begin{equation}
  \ket{\rm final}\simeq W\,e^{i MH \sigma^z}\,W^\dagger 
  \ket{\rm init} = \alpha\ket{N,\ra}+\beta\ket{N{-}2,\la}~.
  \label{e.Wfinal}
\end{equation}
Note that, because of the definition \eqref{e.mapping}, the states
$\ket{N,\ra}$ and $\ket{N{-}2,\la}$ correspond to the right and left output
of the beam splitter in site $N-1$. The derivation presented in this section is 
a rigorous proof of the conjecture made in \cite{kurzynski2011discrete} where 
the authors numerically 
observed that with fully engineered Hamiltonians (i.e. fully engineered coins)
even the {coin degree of freedom is perfectly transferred}. 

The overall scheme is presented in the Supplementary Figs. \ref{fig:pst1} and \ref{fig:pst2}, both for $N$ even
and $N$ odd. A single step moves the photon from an even site to an 
odd site, so an even (odd) number of time steps are required for the
transmission when $N$ is odd (even).
\begin{suppFig}
\begin{center}
  \scalebox{0.3}{
    \begin{tikzpicture}
        \foreach \y in {0,...,1}{
          \bsLm{0}{2*\y+1}{1}{black}{blue}
          \bsRm{0}{2*\y}{Aux}{black}{blue}
          \bsL{1}{2*\y+1}{3}{blue}{blue}
          \bsR{1}{2*\y}{2}{blue}{blue}
          \bsL{2}{2*\y+1}{5}{blue}{blue}
          \bsR{2}{2*\y}{4}{blue}{blue}

          \bsLm{4}{2*\y+1}{N-5}{blue}{blue}
          \bsRm{4}{2*\y}{N-6}{blue}{blue}
          \bsLm{5}{2*\y+1}{N-3}{blue}{blue}
          \bsRm{5}{2*\y}{N-4}{blue}{blue}
          \bsLm{6}{2*\y+1}{N-1}{blue}{black}
          \bsRm{6}{2*\y}{N-2}{blue}{black}
          \inposition{6}{2*\y}{\node at (5.5,4) { \huge\bf Aux };}
        };

        \bsLm{0}{-1}{1}{black}{blue}
        \bsL{1}{-1}{3}{blue}{blue}
        \bsL{2}{-1}{5}{blue}{blue}

        \bsLm{4}{-1}{N-5}{blue}{blue}
        \bsLm{5}{-1}{N-3}{blue}{blue}
        \bsLm{6}{-1}{N-1}{blue}{black}

        \foreach \y in {-1,...,3}{
            \inposition{2}{\y}{\Rbound}
            \inposition{4}{\y}{\Lbound}
        }

        \inposition{0}{4}{
          \node [single arrow, rotate=-90, fill=green!50] at (1,2.5) 
          {\bf \Huge ~ input ~};
          \node [single arrow, rotate=-90, fill=purple!50] at (4,2.5) 
          {\bf \Huge ~ input ~};
        }
        \inposition{6}{-1}{
          \node [single arrow, rotate=-90, fill=green!50] at (4,-2.2) 
          {\bf \Huge ~ output ~};
          \node [single arrow, rotate=-90, fill=purple!50] at (1,-2.2) 
          {\bf \Huge ~ output ~};
        }

        \inposition{3}{4}{\node [scale=2] at (3,2.5) {\bf \Huge $N$ even};}
        \inposition{3}{1}{
          \node [single arrow, scale=1.5, rotate=-90, fill=yellow!30] 
          at (2,2.5) {\bf \Huge odd time steps $M$};}

    \end{tikzpicture}
  }
\end{center}
\caption{Simulation of perfect state transfer with $M$ time steps ($M$ odd). }
\label{fig:pst2}
\end{suppFig}

As one can see from the Supplementary Figs \ref{fig:pst1} and \ref{fig:pst2}, $N-1$ beam splitters 
are required, per line, for 
implementing a CTQW (basically a spin chain dynamics)
of $N$ sites. 
{Two auxiliary sets of beam splitters, with $T_{\rm Aux}=0$,
are required for implementing the 
the open boundary conditions}, though 
the same result can also be obtained with some mirrors in the auxiliary
positions $0$ and $N$. 
\newline


\textbf{Theoretical simulation of state transfer with optimal couplings}.
When minimally engineered systems are considered, only
the ``right coin'' state $\ket\ra$ can be transferred.
Indeed, because of the unitary operator $W$, the left coin state is 
spread between sites $1$ and $2$ before the transmission and, 
unlike fully engineered models,
minimally engineered models are not able to reliably transfer a state from
site $2$ to site $N-1$. For this reason, we consider only the transmission of
the right coin (green light in the the Supplementary Figs \ref{fig:pst1} and
\ref{fig:pst2}).  The transmission quality is measured via
\begin{equation}
  Q = \big\vert\bra{N,\ra} U^M \ket{1,\ra}\big\vert^2,
  \label{e.Q}
\end{equation}
i.e. the probability that a photon goes from $\ket{1,\ra}$
to $\ket{N,\ra}$ after $M$ steps. The results are shown in terms of
    $N$ (horizontal dimension), 
    $M$ (vertical dimension), 
    $T_{\rm bulk}$  ( transmittance of the beam splitters in the ``bulk''
    $T_x =T_{\rm bulk}$ for $x=2,\dots,N-2$), 
    $T_{\rm ends}$ (transmittance of the beam splitters at the ends
    $T_x =T_{\rm ends}$ for $x=1,N-1$) 

\begin{center}
  \label{t.5}
  \bgroup
  \def\arraystretch{1.5}
  \begin{tabular}{|c|c|c|c|}
    \hline
    \multicolumn{4}{|c|}{$N=5$} \\ 
    \hline
    $M$ & $T_{\rm bulk}$ &  $T_{\rm ends}$ & $Q$ \\
    \hline
    6   &   0.919 &     0.750 &     0.886 \\
    8  &   0.673 &     0.500 &     0.921 \\
    10    &   0.484 &     0.345 &     0.942 \\
    12   &   0.358 &     0.250 &     0.956 \\
    14  &   0.273 &     0.188 &     0.965 \\
    16 &   0.214 &     0.146 &     0.972 \\
    18    &   0.172 &     0.117 &     0.977 \\
    20   &   0.141 &     0.095 &     0.981 \\
    22  &   0.117 &     0.079 &     0.984 \\
    \hline
  \end{tabular}
  \hspace{1cm}
  \begin{tabular}{|c|c|c|c|}
    \hline
    \multicolumn{4}{|c|}{$N=6$} \\ 
    \hline
    $M$ & $T_{\rm bulk}$ &  $T_{\rm ends}$ & $Q$ \\
    \hline
    7  &   0.916 &     0.721 &     0.936 \\
    9 &   0.702 &     0.503 &     0.946 \\
    11   &   0.527 &     0.362 &     0.953 \\
    13  &   0.403 &     0.270 &     0.960 \\
    15 &   0.315 &     0.208 &     0.966 \\
    17    &   0.252 &     0.164 &     0.971 \\
    19   &   0.205 &     0.133 &     0.974 \\
    21  &   0.170 &     0.110 &     0.977 \\
    23 &   0.144 &     0.092 &     0.980 \\
    \hline
  \end{tabular}
  \hspace{1cm}
  \begin{tabular}{|c|c|c|c|}
    \hline
    \multicolumn{4}{|c|}{$N=7$} \\ 
    \hline
    $M$ & $T_{\rm bulk}$ &  $T_{\rm ends}$ & $Q$ \\
    \hline
    6   &   0.985 &     0.934 &     0.819 \\
    8  &   0.913 &     0.692 &     0.960 \\
    10    &   0.723 &     0.501 &     0.964 \\
    12   &   0.562 &     0.371 &     0.964 \\
    14  &   0.441 &     0.284 &     0.966 \\
    16 &   0.352 &     0.223 &     0.968 \\
    18    &   0.287 &     0.179 &     0.971 \\
    20   &   0.237 &     0.147 &     0.973 \\
    22  &   0.199 &     0.122 &     0.975 \\
    \hline
  \end{tabular}
  \egroup
\end{center}

Note that for $N=5$ minimally engineered models and fully engineered ones
have the same coupling pattern. In particular, for a fixed transfer time step
$M$, one can evaluate the desired coupling pattern analytically:
$
  T_{\rm ends}  = \sin\left(\frac{2\pi}{M}\right)^2~, 
  T_{\rm bulk}  = \sin\left(\frac{\pi}{M}\sqrt 6\right)^2~. 
  $


\section*{Entanglement growth after a quench} 
We consider one of the most common quenches in the spin chain literature,
namely we focus on the Hamiltonian 
\begin{equation}
  \mathcal H_\Delta = \sum_n j_n (\sigma_n^x\sigma_{n+1}^x + 
  \sigma_n^y\sigma_{n+1}^y +\Delta
  \sigma_n^z\sigma_{n+1}^z )~,
  \label{e.spinchainD}
\end{equation}
and we consider the quench from $\Delta=\infty$ to $\Delta=0$. 
We assume that initially the system is in  the ground state
of $\mathcal H_\infty$, namely  the 
N\'eel state $\ket{\downarrow \uparrow \downarrow \uparrow \downarrow \uparrow \downarrow\dots}$, and then we effectively switch off the 
parameter $\Delta$ so that the system starts to evolve according to the
XY Hamiltonian  $\mathcal H_0$. One can show \cite{alkurtass2014optimal} that
when the couplings $j_n$ 
are engineered for perfect state transfer, then after half the transfer time, namely after $t^*$, 
the initial state $\ket{\downarrow \uparrow \downarrow \uparrow \downarrow \uparrow \downarrow\dots}$ evolves into the state
$\ket{\psi^+_{1,N}}
\ket{\psi^+_{2,N-1}}
\ket{\psi^+_{3,N-2}}\cdots$, where $\ket{\psi^+} \propto \ket{\uparrow \downarrow}+\ket{\downarrow \uparrow}$ and
the subscripts index the spins of the chain. Hence, after half the 
transmission time, two opposite spins lying at the same distance from the
boundaries become maximally entangled. This corresponds to the creation of $N/2$
Bell pairs, namely the maximal amount of pair-wise entanglement. 

In view of the Jordan-Wigner transformation, 
the many-spin anti-ferromagnetic initial state can be simulated using $N/2$
photons in an antisymmetric configuration. Each state $\ket \downarrow$ or $\ket \uparrow$
corresponds respectively to the absence ($\vert 0 \rangle$) or presence ($\vert 1 \rangle$) of a photon. 
We focus on the simplest case, i.e. when $N=5$ and, as a consequence,
the number of particles in the initial state for observing this effect is 
two. Writing the time evolution explicitly for bosons and fermions one can show 
that after half the transmission time 
\begin{equation}
\begin{aligned}
  \ket{\psi(0)}&=a_2^\dagger\,a_4^{\dagger}\ket 0
  &
  \longrightarrow
  \ket{\psi(t^*)} &\propto \begin{cases}
    (a_1^\dagger-a_5^\dagger)(a_2^\dagger+a_4^\dagger)\ket0 
    & \text{for Fermions}, \\\\
    \left[(a_1^\dagger-a_5^\dagger)^2 + (a_2^\dagger+a_4^\dagger)^2
    \right]\ket 0~ & \text{for Bosons} .
  \end{cases}
  \label{e.fermibose}
\end{aligned}
\end{equation}
Therefore, the resulting interference pattern is completely different: while for
bosons the modes are correlated pairwise (namely mode 1 with mode 5, and mode
2 with mode 4), in the fermionic case the resulting state corresponds to a
delocalized particle in mode 1 and 5, and another in modes 2 and 4.
Because of the Jordan-Wigner transformation, the fermionic state in
\eqref{e.fermibose} corresponds to $\ket{\psi^+_{1,5}}
\ket{\psi^+_{2,4}}\ket 0$. 

However, imperfections in the evolution limit the amount of entanglement between the sites. 
To measure the effective generated entanglement we use the entanglement
fraction which, for a two-qubit state, is defined as $\mathcal
E_{ij}=\bra{\psi^+}\rho_{ij}\ket{\psi^+}$, where $\rho$ is a two-qubit density matrix
obtained by tracing all the sites but $i$ and $j$. Because of
\eqref{e.fermibose} we focus on $\mathcal E_{15}$ and $\mathcal E_{24}$. In
order to measure these quantities, we use the entanglement characterisation chip 
to implement the further transformations ${\rm BS}_{15}\,{\rm BS}_{24}$, as
described in the main text, which interfere the distant modes and allow us to
measure entanglement via photodetection in terms of the expectation values 
\begin{equation}
\begin{aligned}
  N'_m &= \langle a_m^\dagger a_m\rangle ~,
 & 
  P'_{nm} &= \langle a_n^\dagger a_m^\dagger a_m a_n\rangle ~.
\end{aligned}
\end{equation}
Indeed, one can show that 
\begin{equation}
\begin{aligned}
  \mathcal E_{15} &= N'_5-P'_{51} ~,
  &
  \mathcal E_{24} &= N'_2 - P'_{24} - P'_{23}+P'_{43}~.
  \label{efrac}
\end{aligned}
\end{equation}
Eqs.\eqref{efrac} give a
practical way of evaluating the simulated entanglement fraction from quantities
that can be measured experimentally.


\section*{Photon Source characterisation}

For this experiment, a mode-locked Ti:Sapphire laser with central wavelength of 785 nm, 160 fs pulse width and 76 MHz repetition rate is used as pump. Second harmonic generation (SHG) in a beta barium borate (BBO) crystal is exploited for conversion into 392.5 nm wavelength. The generation of single-photon pairs is achieved via Spontaneous Parametric Down Conversion (SPDC) by sending the converted pulses through a 2 mm-thick BBO crystal. The generated photons are spectrally filtered by means of interferential filters with 3nm bandwidth, centered at 785 nm. Typical detection rates of generated photons were approximately $\sim 120 $ KHz for singles and $\sim 7$ KHz for pairs. Schematics of the source is shown in Supplementary Fig. \ref{sorgente}. The unconverted residual beam at 785 nm from the SHG is separated through a dichroic mirror and used as simulation beam for alignment and classical characterisation of the integrated device. 

The generated photon state is in the general entangled form $\ket{\Psi^{\chi}_{2 \mathrm{p}}} = \frac{\ket{HV} + e^{\imath \chi}\ket{VH}}{\sqrt{2}}$, where the phase $\chi$ depends on the angle of the crystal's optical axis. Due to birefringence and dispersion effects in the SPDC BBO crystal, a walk-off compensation between the two polarizations is required for full indistinguishability. This is done by using a combination of an half-wave plate and a 1mm thick BBO crystal in each of the two photons' paths. 

A controlled liquid crystal retarder is used in one arm to vary the phase $\chi$, allowing to attain the required states. Half-wave plates and polarizing beamsplitters at each arm can be inserted when necessary to perform polarization analysis. The photons are then collected into single-mode optical fibers mounted on 3-paddle polarization controllers, used to compensate the polarization due to the bending in the fibers, and sent through delay lines to achieve temporal synchronization of the photons. Indistinguishability of the photon source is measured by performing an HOM \cite{HOM} interference experiment in a 50/50 beam splitter (BS) after polarization analysis. The obtained visibilities are $V^{\mathrm{source}}_{\mathrm{raw}} = - 0.914 \pm 0.002$ for raw data and $V^{\mathrm{source}}_{\mathrm{corr}} = - 0.95 \pm 0.01$ corrected for accidental counts (see Supplementary Fig \ref{dippeak}a). Similar interference measurements have been performed for the two entangled states $\ket{\Psi^{+}_{2 \mathrm{p}}}$ and $\ket{\Psi^{-}_{2 \mathrm{p}}}$ to characterize the quality of the generated entangled. The observed visibilities where $\textit{V}^{\mathrm{B}}_{\mathrm{raw}} = - 0.880 \pm 0.002$ and $\textit{V}^{\mathrm{F}}_{\mathrm{raw}} = 0.90\pm0.01$ for raw measurements, and $\textit{V}^{\mathrm{B}}_{\mathrm{corr}} = - 0.957 \pm 0.002$ and $\textit{V}^{\mathrm{F}}_{\mathrm{corr}} = 0.97 \pm 0.01$ corrected for accidental counts (see Supplementary Fig \ref{dippeak}b).

\begin{suppFig}[h!]
  \centering
    \includegraphics[width=0.8\textwidth]{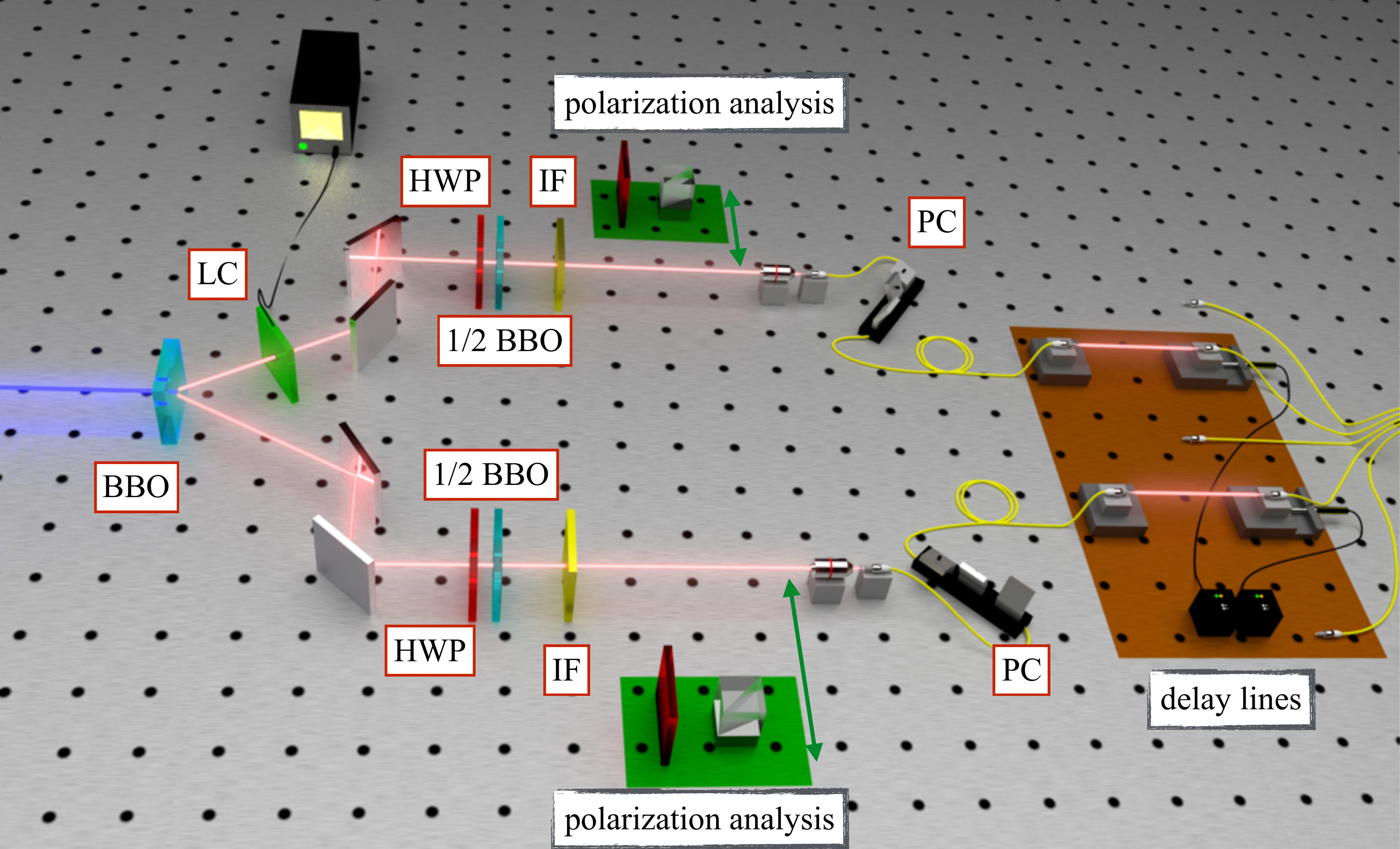}
 \caption{\footnotesize \textbf Visual schematics of the single-photon source used for the experiment. BBO: beta barium borate crystal used for parametric down conversion; LC: liquid crystal for entanglement phase $\chi$ control; HWP \& 1/2 BBO: half-waveplate and half beta barium borate crystal for walk-off compensation; PC: polarization compensation. The polarization analysis is comprised of half-waveplate and polarizing beam splitter. The delay line is used to temporally synchronize the photon and couple them into the single mode fiber array.}
\label{sorgente}
\end{suppFig}

\begin{suppFig}[h!]
  \centering
    \includegraphics[width=1\textwidth]{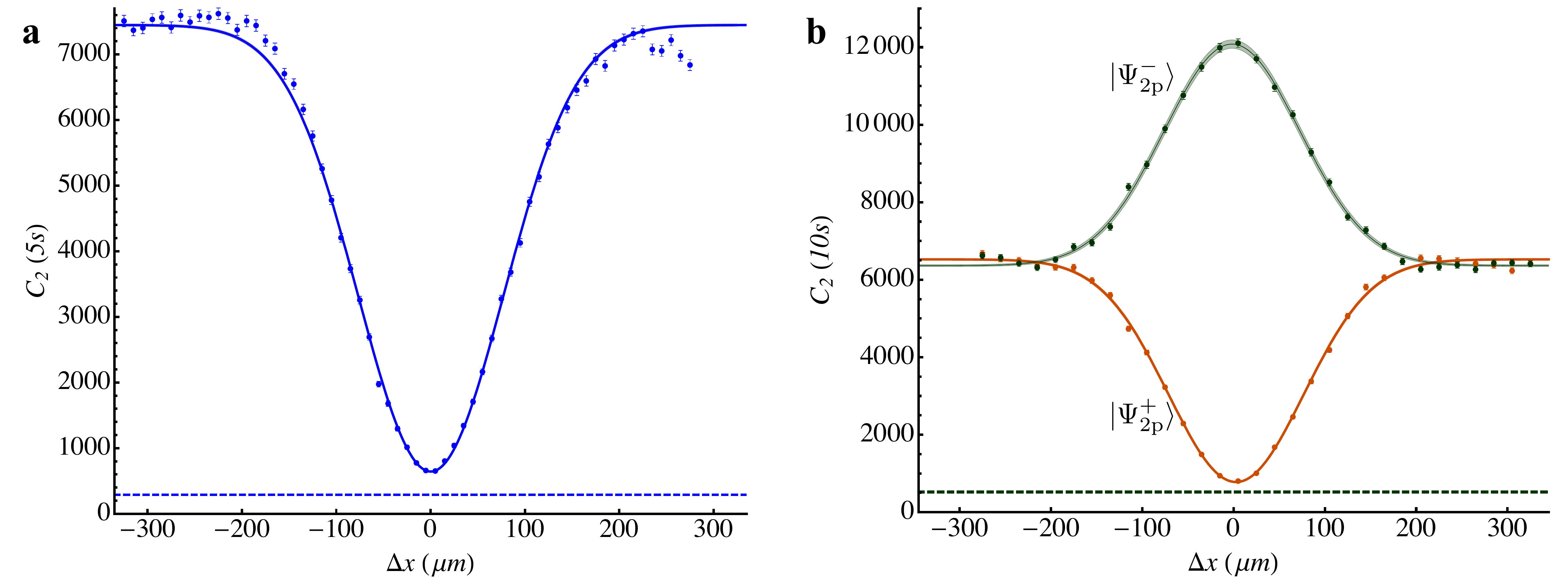}
 \caption{\footnotesize \textbf{a.} Hong-Ou-Mandel interference scan over 600 $\mu$m of two indistinguishable photons. \textbf{b.} Hong-Ou-Mandel interference dip for state $\vert \Psi^{+}_{2 \mathrm{p}}\rangle$ (green) and peak for state $\vert \Psi^{-}_{2 \mathrm{p}}\rangle$ (orange).} 
 \label{dippeak}
\end{suppFig}


\section*{Characterisation of the QTC device}

The generated photons are coupled into the integrated device using a single-mode fiber array, and collected after evolution through the interferometer using a multimode fiber array which sends the photons to Single Photon Avalanche Photodiodes (SPADs). Both input and output fiber arrays are mounted on roto-translational stages. The alignment is performed using the unconverted residual from the SHG. Overall transmission efficiency from delay lines to the photodetectors with the first chip is approximately 12\%, which includes transmission losses inside the chip and coupling losses at the three interfaces between delay line, input fiber array, chip and output fiber array. The phase $\chi$ of the entangled state is set by exploiting HOM interference for the output bunching contributions. This is obtained by inserting an additional in-fiber BS at output mode 1 and by measuring the output two-fold coincidences. By maximizing the coincidence rate we have a symmetric (bosonic) state $\ket{\Psi^{+}_{2\mathrm{p}}}$ and by minimizing it we have an anti-symmetric (fermionic) state $\ket{\Psi^{-}_{2\mathrm{p}}}$.

Photon pairs are sent into the device to simulate bosonic and fermionic transport in the lattice, respectively exploiting the symmetric and anti-symmetric nature of the Bell states $\ket{\Psi^{+}_{2 \mathrm{p}}}$ and $\ket{\Psi^{-}_{2 \mathrm{p}}}$. The simulation of the transport of a spin state in a 1D chain, initially in a N\'{e}el's state $\ket{\downarrow \uparrow \downarrow \uparrow \downarrow}$, is obtained by injecting the $\ket{\Psi^{-}_{2\mathrm{p}}}$ state in input 2 and 4. Switching from one state to the other is achieved by changing the phase $\chi$ with the liquid crystal retarder. Detection of output states in which two photons are in the same spatial mode is obtained by inserting an in-fiber 50/50 beam splitter to the chip's output.
\newline

\textbf{Tomography of the QTC device}.
The reconstruction of the unitary matrix which describes the actual interferometer corresponds to retrieving the values of its elements (moduli and complex phases). Unitary's moduli are sufficient to describe single-particle and classical behavior, while phase terms are crucial for two-particle interference. The unitary matrix is experimentally reconstructed by measuring the single-photon output distributions (or the equivalent power splitting ratios with classical light) for all inputs and two photon interference for several pairs of inputs \cite{Laing12,Crespi15}. More specifically, we measured the power splitting ratios with classical light and two-photon Hong-Ou-Mandel visibilities for $6$ different input states for both polarizations \textit{H} and \textit{V} (see Supplementary Fig. \ref{data_tomography}).
Characterisation of the device is performed starting from the chip structure of Fig. 1b and by assuming unknown values for all the fabrication parameters (directional couplers trasmittivities and phases between the modes). The value of the parameters are obtained by minimizing a suitable $\chi$-square function, while errors in the reconstruction are retrieved with a Monte Carlo simulation starting from the experimental power-splitting ratios and two-photon data \cite{Crespi15}.

\begin{suppFig}[t!]
  \centering
    \includegraphics[width=1\textwidth]{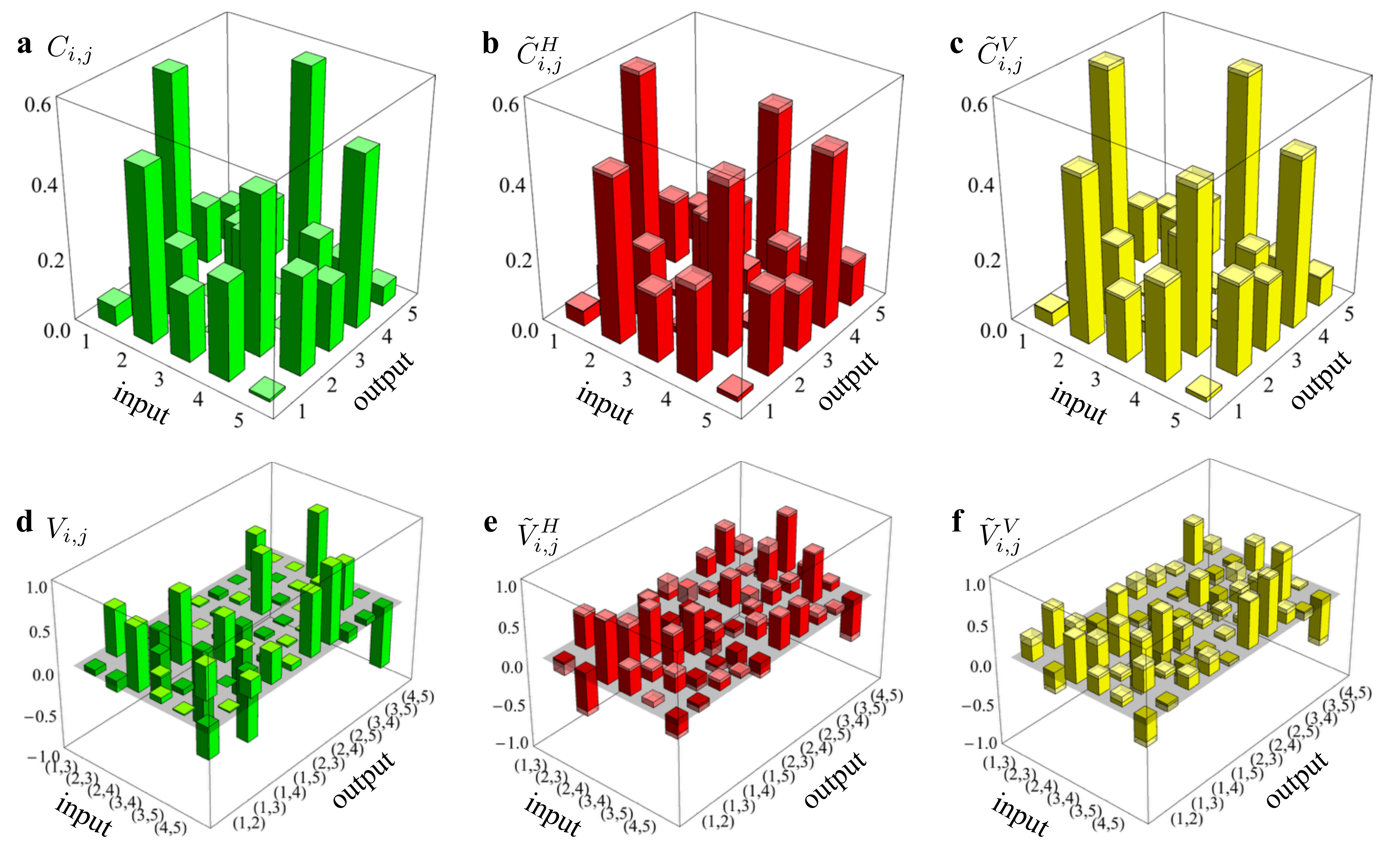}
 \caption{Power-splitting ratios with classical light (\textbf{a-c}) and two-photon interference dips and peaks visibilities (\textbf{d-f}) results: \textbf{a-d} calculations from the theoretical unitary $U$, \textbf{b-e} experimental results with horizontal polarization and \textbf{c-f} experimental results with vertical polarization. Lighter regions are $1\sigma$ experimental errors.} 
\label{data_tomography}
\end{suppFig}

The obtained fidelities for the two polarizations with respect to the theoretical model are $\mathcal{F}^{H} = 0.962\pm0.001$ and $\mathcal{F}^{V} = 0.977\pm0.002$. The polarization insensitivity of the device is confirmed by the results for the fidelity between the reconstructed unitary in \textit{H} and \textit{V} polarization of $\mathcal{F}^{H/V}_{\mathrm{Reco}} = 0.99\pm0.01$ (see Supplementary Fig. \ref{tomography_chip}). Experimental single- and two-photon distributions, measured in both polarizations \textit{H} and \textit{V}, further confirm the polarization insensitivity (see Supplementary Fig. \ref{Measurements_SI}).

\begin{suppFig}[t!]
  \centering
    \includegraphics[width=1\textwidth]{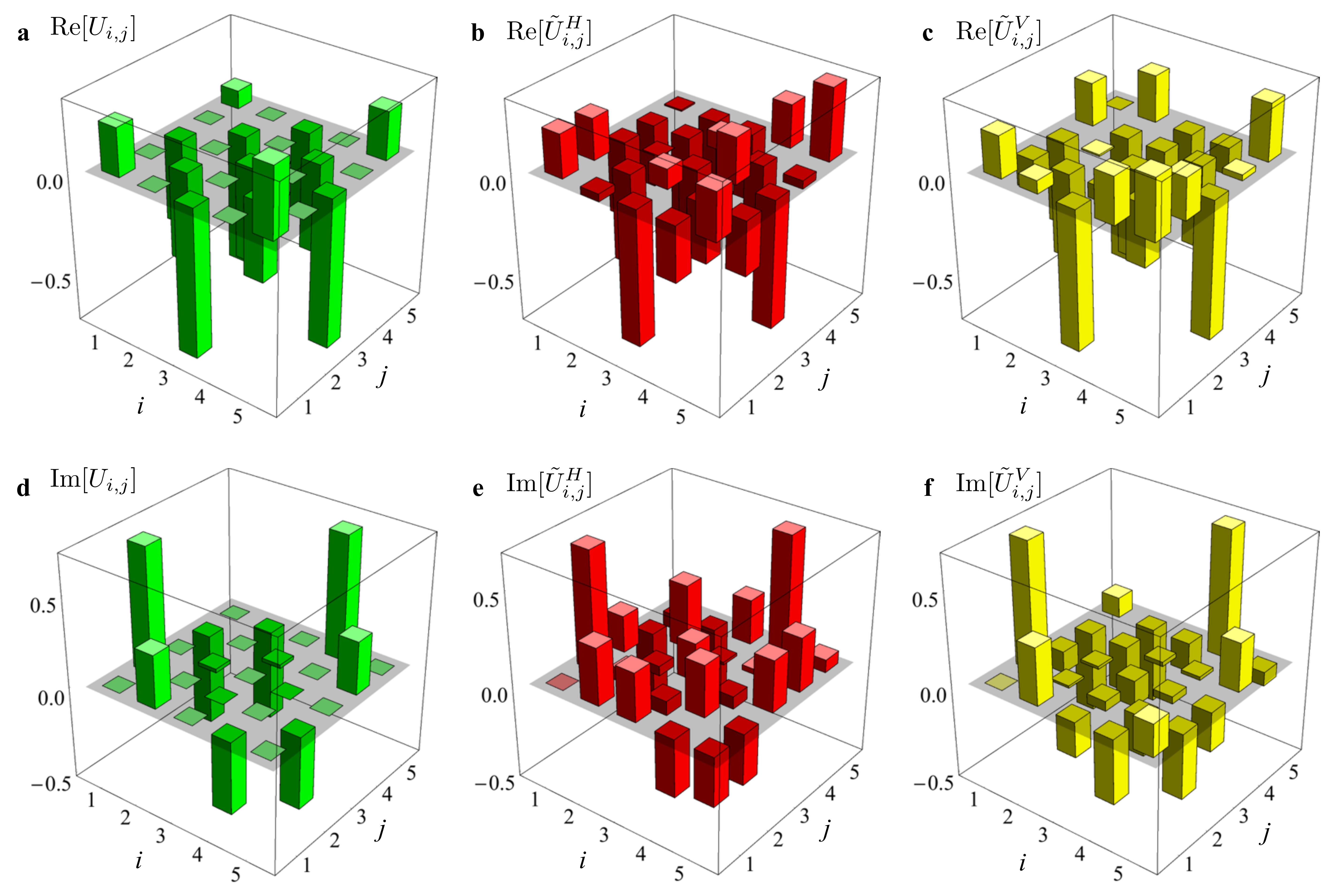}
 \caption{Real and imaginary terms for \textbf{a-d}  theoretical unitary matrix, \textbf{b-e} reconstructed unitary matrix for horizontal polarization, and \textbf{c-f} reconstructed unitary matrix for vertical polarization.} 
\label{tomography_chip}
\end{suppFig}

\begin{suppFig}[t!]
  \centering
    \includegraphics[width=1\textwidth]{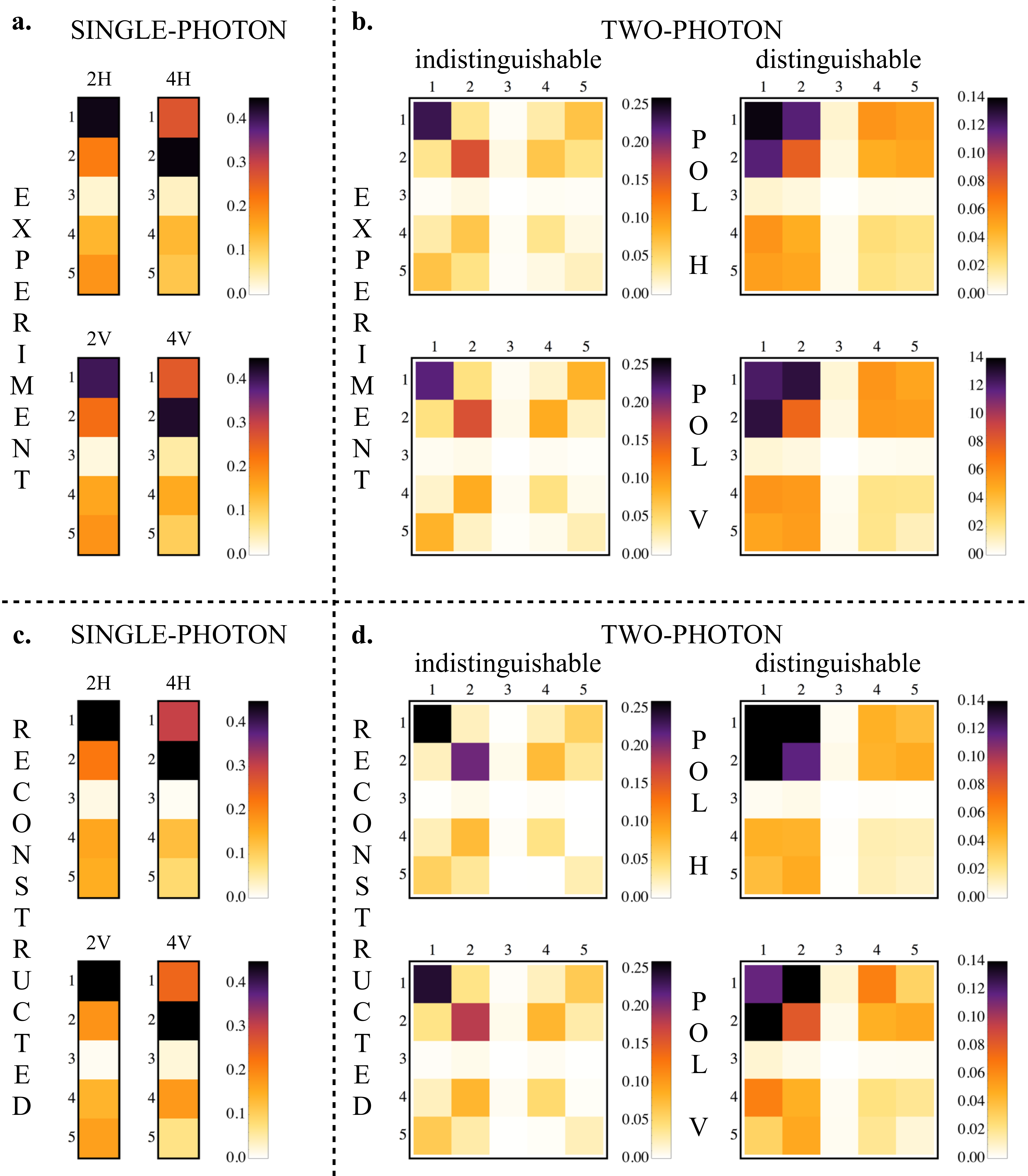}
 \caption{\textbf{a-b.} Experimental single-photon couplings for input 2 and 4 and two photon unitary transformation, both indistinguishable and distinguishable, for polarization $H$ and $V$. \textbf{c-d.} Numerical prediction from experimentally reconstructed unitary transformation of single-photon probabilities for input 2 and 4 and two-photon interference, both indistinguishable and distinguishable, distribution for polarization $H$ and $V$. Both single- and two-photon distributions show the polarization insensitivity of the device.} 
 \label{Measurements_SI}
\end{suppFig}


\section*{Modeling the evolution with the reconstructed unitary transformation}

Here we briefly discuss how to model the evolution of a polarization entangled state through the quantum transport interferometer. We consider the evolution of an entangled pair through a unitary operator $U^{\pi}$, which in this case presents a small difference for the two orthogonal polarizations. The action of $U^{\pi}$ on the field operator before ($a_{j,\pi}^{\dag}$) and after ($b_{i,\pi}^{\dag}$) the evolution, where $\pi=H,V$ labels the polarization state, is expressed by $a_{j,\pi}^{\dag} = \sum_{i} U^{\pi}_{ji} b_{i,\pi}^{\dag}$. The input state of the process is a two-photon entangled pair in the $\vert \Psi^{\pm}_{2 \mathrm{p}} \rangle_{i,j}$ Bell state, where the photons are injected in input ports $i$ and $j$:
\begin{equation}
\vert \Psi^{\pm}_{2 \mathrm{p}} \rangle_{i,j} = \frac{1}{\sqrt{2}} (\vert H \rangle_{i} \vert V \rangle_{i} \pm \vert V \rangle_{i} \vert H \rangle_{j})
= \frac{1}{\sqrt{2}} (a_{i,H}^{\dag} a_{j,V}^{\dag} \pm a_{i,V}^{\dag} a_{j,H}^{\dag}) \vert 0 \rangle.
\end{equation}
The state after the evolution can be written as:
\begin{equation}
\begin{aligned}
\vert \Psi^{\pm}_{2 \mathrm{p}} \rangle_{i,j} &\stackrel{U_{\pi}}{\rightarrow} \frac{1}{\sqrt{2}} \left[ \left( \sum_{m} U^{H}_{i,m} b_{m,H}^{\dag} \right) \left( \sum_{n} U^{V}_{j,n} b_{n,V}^{\dag} \right) \pm \left( \sum_{n} U^{V}_{i,n} b_{n,V}^{\dag} \right) \left( \sum_{m} U^{H}_{j,m} b_{m,H}^{\dag} \right) \right] \vert 0 \rangle=\\
&= \frac{1}{\sqrt{2}} \left[ \sum_{m,n} \left(U^{H}_{i,m} U^{V}_{j,n} \pm U^{H}_{j,m} U^{V}_{i,n} \right) b_{m,H}^{\dag} b_{n,V}^{\dag} \right] \vert 0 \rangle.
\end{aligned}
\end{equation}
The transition amplitudes from input state $\vert \Psi^{\pm}_{2 \mathrm{p}} \rangle_{i,j} $ to the output configurations $\vert rH, sV \rangle$ and $\vert rV, sH \rangle$ for $r \neq s$ are respectively:
\begin{eqnarray}
\langle r H, s V \vert U \vert \Psi^{\pm}_{2 \mathrm{p}} \rangle_{i,j} &=& \frac{1}{\sqrt{2}} \left( U^{H}_{i,r} U^{V}_{j,s} \pm U^{H}_{j,r} U^{V}_{i,s} \right) \\
\addtocounter{equation}{1}
\langle s H, r V \vert U \vert \Psi^{\pm}_{2 \mathrm{p}} \rangle_{i,j} &=& \frac{1}{\sqrt{2}} \left( U^{H}_{i,s} U^{V}_{j,r} \pm U^{H}_{j,s} U^{V}_{i,r} \right)
\end{eqnarray}
where the $\pm$ signs in the transition amplitudes depend on the symmetry of the input entangled state. We can define the submatrices $U^{i,j}_{r,l}$ of $U^{H}$ and $U^{V}$ as:
\begin{equation}
\label{eq:Uijrs}
U^{i,j}_{r,s} = \begin{pmatrix} U^{H}_{i,r} & U^{H}_{j,r}\\ U^{V}_{i,s} & U^{V}_{j,s} \end{pmatrix}.
\end{equation} 
If the input state is in the singlet anti-symmetric state $\vert \Psi^{-}_{2 \mathrm{p}} \rangle_{i,j}$, the probability of obtaining a photon on modes $r$ and $s$ can be expressed as:
\begin{eqnarray}
\mathrm{Prob}(r,s \vert \Psi^{-}_{i,j}) &=&\frac{1}{2} \left[ \left \vert \det \left(U^{i,j}_{r,s}\right)\right\vert^{2} + \left\vert \det \left(U^{i,j}_{s,r}\right)\right\vert^{2} \right], \\
\addtocounter{equation}{1}
\mathrm{Prob}(r,r \vert \Psi^{-}_{i,j}) &=& \frac{1}{2} \left \vert \det \left(U^{i,j}_{r,r}\right)\right\vert^{2}
\end{eqnarray}
thus depending on the determinants of the submatrices $U^{i,j}_{r,s}$. Conversely, if the input state is symmetric $\vert \Psi^{+}_{2 \mathrm{p}} \rangle_{i,j}$ we obtain:
\begin{eqnarray}
\mathrm{Prob}(r,s \vert \Psi^{-}_{i,j}) &=&\frac{1}{2} \left[ \left \vert \per \left(U^{i,j}_{r,s}\right)\right\vert^{2} + \left\vert \per \left(U^{i,j}_{s,r}\right)\right\vert^{2} \right], \\
\addtocounter{equation}{1}
\mathrm{Prob}(r,r \vert \Psi^{-}_{i,j}) &=& \frac{1}{2} \left \vert \per \left(U^{i,j}_{r,r}\right)\right\vert^{2}
\end{eqnarray}
thus depending on permanents of the submatrices $U^{i,j}_{r,s}$.


\section*{Characterisation of the ECC device}
For an ideal fermionic quantum spin transport, the target state would be of the form $\ket{\psi_{\mathrm{ideal}}} = \ket{\psi^{+}_{1\mathrm{p}}}_{24}\ket{\psi^{+}_{1\mathrm{p}}}_{15}\ket{0}_{3}$. Our device is a finite engineered chain and the generated the state is $\ket{\psi_{\mathrm{out}}} = (\alpha\ket{10}_{15}+\beta\ket{01}_{15}) (\gamma\ket{10}_{24}+\delta\ket{01}_{24})\ket{0}_3$, where the terms $\alpha$, $\beta$, $\gamma$ and $\delta$ define an unbalancement due to the approximation of the perfect state.

The coherence between the path entangled states is studied using the second device, comprised of integrated beam splitters and thermal phase shifters. In particular, with suitable phases before the two beam-splitters, the target state would result in a photon in output mode 2 and a photon in output mode 5. The phase $\chi$ of the liquid crystal for the injection of a singlet $\ket{\Psi^{-}_{2\mathrm{p}}}$ state is set by measuring the physical quantity $(P'_{\mathrm{12}} + P'_{\mathrm{14}} + P'_{\mathrm{25}} + P'_\mathrm{{45}}) - (P'_{\mathrm{24}} + P'_{\mathrm{15}})$ at the output of the second device, where $P'_{ij}$ is the coincidence probability at modes \textit{i} and \textit{j}. Such quantity is maximized for input state $\ket{\Psi^{-}_{2 \mathrm{p}}}$ and minimized for input state $\ket{\Psi^{+}_{2\mathrm{p}}}$.

The phase shifters in the second device are actively controlled via thermal resistors. Each thermal resistor is connected to an external power supply with independent voltage channel control and the applied voltage ranges between 0 V to 7 V, in which a full $2\pi$ oscillation period is seen. Overall transmission efficiency from delay lines to the photodetectors is approximately 0.6\%, which includes transmission losses inside the two chips and coupling losses at the four interfaces between delay line, input fiber array, first chip, second chip and output fiber array. Coincidence rates for all outputs are measured by singularly tuning the applied phase shift from each thermal resistors, showing a $2\pi$ oscillation within $\sim 0.8$ W dissipated power. When evaluating the expected curves for the quantities $S_1$ and $S_5$ as a function of $\phi_5$, losses corresponding to an additional efficiency $\eta_5=0.36$ is considered for mode 5 at the interface between the two devices. This quantity is retrieved from characterisation with classical light.
Each experimental distribution is compared to its theoretical prediction though the measurement of the similarity, which is defined in terms of all contributions of the correlation function

\begin{equation}
 \mathcal{S}^{\mathrm{B(F)}} = \frac{\left(  \sum_{ij}  \sqrt{\Gamma_{ij}^{\mathrm{B(F)}}  \tilde{\Gamma}_{ij}^{\mathrm{B(F)}}} \right) ^{2} }{ \left(  \sum_{ij}  \Gamma_{ij}^{\mathrm{B(F)}}  \sum_{ij}  \tilde{\Gamma}_{ij}^{\mathrm{B(F)}}  \right)}. 
 \label{eq:similarity}
 \end{equation}
where the $\tilde{\Gamma}_{ij}$ and $\Gamma_{ij}$ are experimental and theoretical correlation function for mode \textit{i} and \textit{j} for bosons (B) and fermions (F).

\begin{suppFig}[h!]
  \centering
    \includegraphics[width=0.8\textwidth]{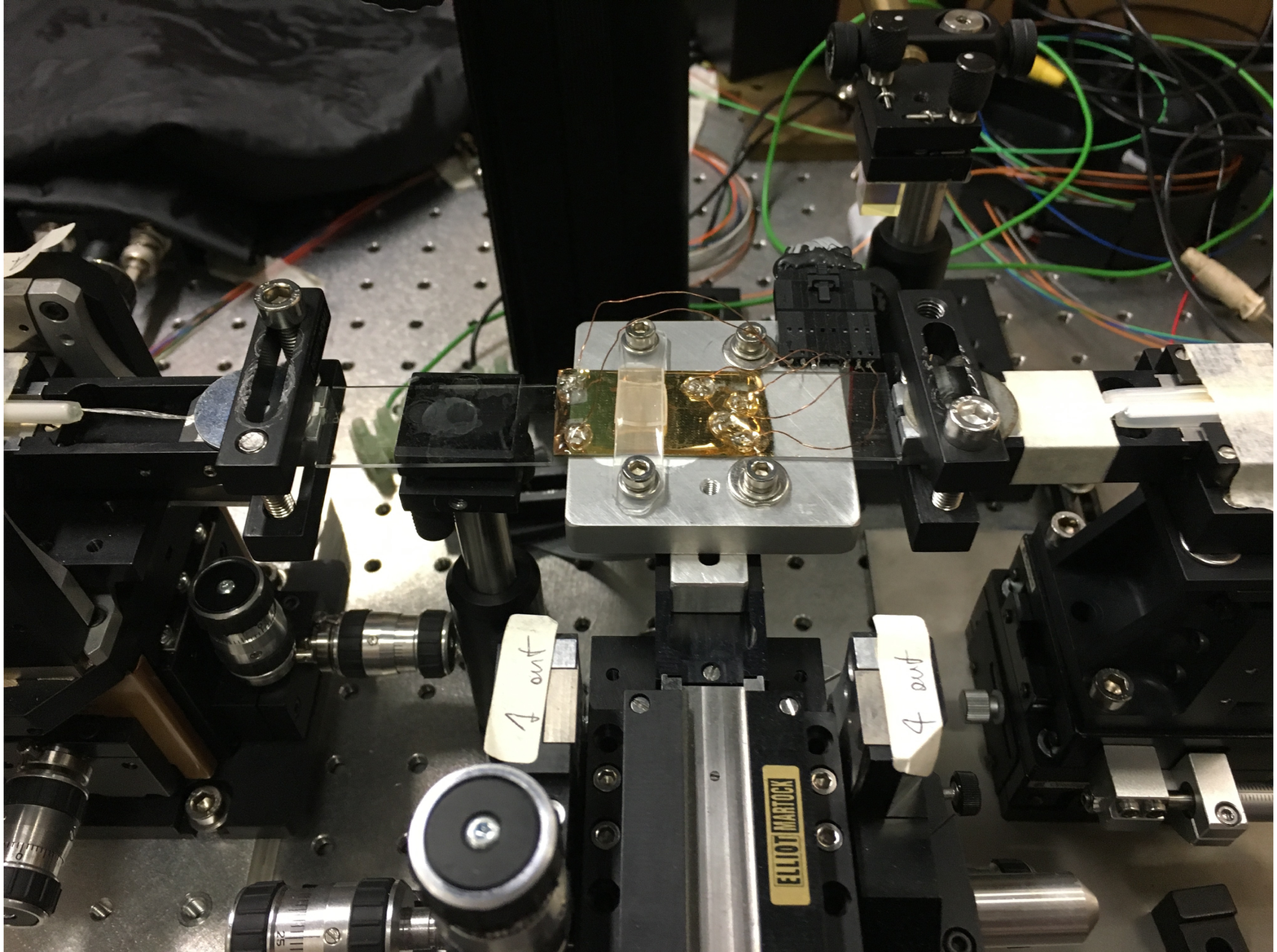}
 \caption{Photograph of the experimental setup. From the left: input single-mode fiber array, chip with quantum transport device, chip with thermal resistor and entanglement check device, output multi-mode fiber array.} 
 \label{fotochip}
\end{suppFig}